\documentclass{jaa} 
\usepackage{natbib}
\usepackage[modulo]{lineno}

\usepackage{graphicx}

\usepackage{xcolor}
\begin{document}\sloppy

\title{Testing the Simultaneity of Forbush Decreases with \\Algorithm-Selected Forbush Event Catalogue}


\author{J. A. Alhassan\textsuperscript{1,*}, O. Okike\textsuperscript{2} \and A. E. Chukwude\textsuperscript{1}}
\affilOne{\textsuperscript{1}Department of Physics and Astronomy, University of Nigeria, Nsukka, Nigeria.\\}
\affilTwo{\textsuperscript{2}Department of Industrial Physics, Ebonyi State University, Abakaliki, Nigeria.}


\twocolumn[{

\maketitle

\corres{jibrin.alhassan@unn.edu.ng}


\begin{abstract}

Accurate detection and precise timing of transient events such as X-ray photons, $\gamma$-ray burst, coronal mass ejections (CMEs), ground level enhancements (GLEs) and Forbush decreases (FDs) frequently raise issues that remain on the cutting edge of research in astrophysics. In an attempt to automate FD event selection, a combination of Fast Fourier transform as well as FD detection algorithms implemented in the statistical computing software R was developed and recently used to calculate the magnitude and FD event timing.
The R-FD code implemented in the present study includes several different calculations. Some subroutines detect both small and large transient intensity reductions (minima/pits) as well as increases (maxima/peaks) in cosmic ray (CR) data. Others calculate event amplitude, timing and cataloging of the events identified. As the current work focuses on reductions in CR flux (FDs), the subroutine that identifies increases was disabled. Totals of 229 FDs at Magadan neutron monitor (NM), 230 (Oulu NM) and 224 (Inuvick NM) were identified with daily averaged data, while 4032 (Magadan), 4144 (Oulu) and 4055 (Inuvick) were detected with hourly averages. FDs identified as simultaneous at the three stations totaled 99 for the daily and 261 for the hourly CR averages respectively.


 \end{abstract}

}]



\doinum{12.3456/s78910-011-012-3}
\volnum{000}
\year{0000}
\pgrange{1--}
\setcounter{page}{1}
\lp{1}

\section{Introduction}
Galactic cosmic rays (GCRs) are cosmic rays that have their origin inside our Galaxy. They are high-energy charged particles, usually protons, electrons, and fully ionized nuclei of light elements. GCRs are ubiquitous in the interplanatary (IP) space \citep{pap:2019}. Periodically and abruptly, GCR flux  experiences variabilities which are attributed to solar wind interplanetary magnetic field (IMF) structures such as shocks, sheaths, coronal mass ejections/interplanetary coronal mass ejections (CMEs/ICMEs), magnetic clouds (MCs) and  corotating interaction region (CIR)  \citep[e.g.][] {sven:2012}. The periodic long term  modulation of GCR intensity  include: diurnal, 27-d, and 11-year modulations. The non-periodic short term variabilities  are Forbush decreases (FDs)  and ground level enhancements (GLEs).

FDs are generally regarded as the  short-term non-periodic reductions in the  GCR flux  measured with neutron monitors (NMs) at Earth \citep{do:63,oki:20}.  Scott Forbush was the pioneer investigator who discovered the depression in cosmic ray flux while working with ionization chamber about eight decades ago  \citep{fo:37-2}. Typically, FDs have well-defined profiles made up of four parts \citep[e.g.][]{oh:08}: onset, main phase, point of maximal depression and recovery phase. The entire FD profile  can last on a time scale from several hours to several days \citep{pap:2010}.

Investigations relating to  simultaneous observations of FD events by different ground level NMs situated globally  on the Earth's surface remains a subject of  interest \citep[e.g.][]{Barrantes:2018}. Forbush decrease events are expected to occur simultaneously and globally since GCRs are uniformly distributed  in the IP space. However, empirical evidences  of non-simultaneity have also been reported \citep[e.g.][]{oh:08, pap:2019}. 

 \citet{oh:08} investigated the  hourly data of the Oulu neutron monitor (NM) station from 1998 to 2002. From the  49 FD events, with a pre-defined threshold of  $>$ 3.5\% flux reductions they identified, global simultaneity was
determined by comparing the time profiles of these FD events with those recorded  at Inuvik and Magadan NM stations. The FD event is defined to be simultaneous if the main phases of the GCR intensity time variation profiles overlap in universal time (UT)  while non-simultaneous event refers to when the main phases of the FD event  overlap in local time (LT) irrespective of the station's location. They  argued that the occurrence of simultaneous FD events and non-simultaneous FD events is dependent on the intensity and propagation direction of MCs. In other words, simultaneous FD events are observed when IP shocks and MCs followed by strong magnetic fields head directly for the Earth, and are distributed symmetrically, so that the Earth is at the center of the magnetic barrier. Thirty seven out of 49 FDs were detected by the three separate stations simultaneously in UT, while the rest 12 of the FD events were detected simultaneously at the same LT. They noted that large amplitude events, on average, were simultaneously observed in  UT  while  small amplitude  events are generally non-simultaneous.

In a follow up work, \citep{oh:09}, selected 218 FD events which had a threshold of $>$3.0\% GCR
intensity decrease from  the hourly data of Oulu, Inuvik and Magadan   that span a period of 36 years from 1971 to 2006.They reported  that while 165 out of 218 FD events were detected by the three separate stations simultaneously in UT, 53 events were detected simultaneously at the same LT. 

\citet{le:2015} selected 220 FD events with a cosmic ray CR intensity
variation of at least 3.0\% from  three middle-latitude NM stations (Climax, Irkutsk and Jungfraujoch) from 1971 to 2006. The authors observed that out of the 220 FD events selected during the 36 year period, the main phase of 167 FD events were overlapped in UT suggesting that they are  simultaneous. On the other hand, the main phases of the remaining 53 FD events are overlapped in  LT which was interpreted as indication that the events are  non-simultaneous. 

The method of FD identification has been predominantly manual for the past eight decades \citep{ok:2020}. Manual technique   involves  downloading of CR data of preferred resolution, visually searching for points of maximum depression from the plotted data, noting the main phase onset and end time dates  and calculating the individual FD amplitude. This method has been used by several authors \citep[e.g.][]{oh:08, kris:08, oh:09, le:2015} and references therein. 
The  manual method  is laborious and deficient in several ways. The limitations ranges from its inability to account for unwanted signals in the CR data such as diurnal CR anisotropies, limited number of FDs, bias and subjectivity arising from trial and error to the problem of being able to simultaneously analyze a good number of  FDs at many NM stations  \citep[e.g.][]{oki_:2020}.

Currently, only IZMIRAN group and a few other investigators  \citep[e.g.][]{rami:2013,ok:2019, Light:2020} have used automated technique in FD detection.  The IZMIRAN group  employed  the global survey method (GSM) in which they assimilated CR data from an array of all  NMs  in their FD detection.

 \citet{rami:2013} developed an interface definition language (IDL) protocol in an attempt to identify FDs. They asserted  that FD is the most spectacular variability in the GCR intensity, hence its  selection requires sophisticated technique rather than the prevailing manual method. With their automated program, they  obtained a catalogue of FDs from CR data of the major part of solar cycle 23 (1996- 2008).

\citet{ok:2019} developed a Fourier transform technique and R algorithm to calculate the FD amplitude and time of minimum depression in the GCR intensity. This automated technique is designed to  filter out both the low -and high-frequency superposed signals using Fourier decomposition. The R code then searches for the points of minimum depression and calculates their FD event magnitudes and time of event in the high-frquency data. This FD location algorithm was modified as  \citep{ok:2020}. Here, the program  accept CR raw data as the input signal instead of the high-frequency signal. The second code tracks the peaks/maxima or pits/minima and also  their time of occurrence in CR data.

\citet{oki_:2020} further developed an algorithm used to identify FDs from unprocessed and Fourier transformed CR data. This code accepts raw CR data as input signal. The author employed Fourier transform technique (FTT) and an R program in order to determine the FD magnitude and the corresponding time of minimum depression in the GCR intensity. The  FTT algorithm is aimed at   transforming CR signal into sine and cosine waves. The essence of this is to arrive at a signal which is easier to use than the original one. 

\citet{Light:2020} determined FD amplitude  using an expression in the form of  equation 1. They used a 24-hour average to estimate the magnitude of each of the FD events. As good and new as this approach may appear, it might be  tedious and time-consuming as in the case of the manual technique.   In this paper, we seek to address  the simultaneity of FDs by employing computer software-selected FD  sample. 

\subsection{Motivation for the Current Work}

Simultaneity of FD event is well documented in  the literature. However, a survey of the literature reveal that the methods adopted by several researchers in confirming event simultaneity is largely not clear. \citet{oh:08} reported that 37 FD events out of the 49 events that satisfy their predefined baseline condition were simultaneous. In the last column of their Table 1, simultaneity was confirmed with a  ''YES" or ''NO" remark. \citet{lee:2013} in like manner, showed in their Table 3 a ''YES" or ''NO" indicator for the 35 simultaneous FDs out of the 45 events they selected. No further detail on how they arrived at this vital conclusion was given. 

\citet{oh:09}  selected 218 events that appear as FDs at each NMs. From this catalogue, they declared 165 FDs simultaneous while 53 events were non-simultaneous. Even though the criteria of comparing the time profiles of the simultaneous FD candidates of one stations with those of other stations was documented, the details of the simultaneous FD selection  is not explicit. \citet{kris:08}  identified  FDs employing Climax neutron monitor record as a basis. In their attempt to identify simultaneous events, the FDs at Climax were  compared to FDs found at OULU and Moscow NM stations. No more clear detail on this was reported.   

 We feel that even though the publications of simultaneous FDs in the literature have contributed substantially to our understanding of this phenomenon, a clear and an unambiguous method of arriving at FDs that are simultaneous or non-simultaneous ought to be provided. We demonstrate explicitly, simultaneity of FDs using computer algorithm-selected FD catalogue in this paper.

\section{Data}
In this submission, we obtained the raw  CR data in the same format, pressure corrected at different resolutions from IZMIRAN common website : http://cr0.izmiran.ru/common. The site is hosted by the Pushkov Institute of Terrestrial Magnetism, Ionosphere, and Radio Wave Propagation, Russian Academy of Sciences (IZMIRAN) team.  

Our investigation is based on daily averaged CR data from Magadan (MGDN), Oulu (OULU) and Inuvik (INVK) NMs covering the period 1998 to 2002. The parameters of these NMs are presented in Table 1. 
 

\begin{table}[ht]
\caption{Parameters of MGDN, OULU and INVK NMs taken from \citep{oki_:2020}. "Lon", "Lat", "R", and "Alt" respectively stand for longitude, latitude, effective vertical cutoff rigidity, and altitude.  }
\centering
\begin{tabular}{c c c c c c}
  \hline\hline
S/N & NMs & Lon($^{\circ}$) & Lat($^{\circ}$) & R(GV) & Alt(m) \\ [0.5ex]
\hline
1 & MGDN & 151.02E & 66.12N & 1.99 & 220\\
2 & OULU & 25.47E & 65.05N & 0.77 & 0\\
3 & INVK & -133.72W & 63.35N & 0.17 & 21\\ [1ex]
\hline
\end{tabular}
\end{table} 


\section{Illustration of Simultaneous and Non-simultaneous FDs}

Researchers that investigate simultaneity and non-simultaneity of FD events define the concepts variously. \citet{oh:08, oh:09, lee:2013, le:2015} defined simultaneity of FDs with regards to overlap of FD main phase in UT while non-simultaneity FDs  refer to the separation of the main phase in LT. \citet{ok:2011} used the complete FD event profile (onset, main phase, point of maximum depression and the recovery phase) to define FD event simultaneity. Accordingly, a significant overlap among the FD event profiles at all the NMs, their location at the Earth not withstanding, is regarded as simultaneous FDs. In this paper, simultaneous and non-simultaneous FD events are defined with respect to the time of FD event point of maximum reduction. Hence simultaneous FDs refer to events whose points of maximum reduction are detected at the same time by different NMs. Non-simultaneous FDs, on the other hand, are those whose points of maximum  CR reduction occured at different times  at different NM stations.

As an example, CR intensity profiles of May 23, 2002 is plotted in Figure 1 for Climax, Irkutsk, Inuvik, Oulu and Magadan NM stations. The maximum depression of the CR counts for the five stations occured on  May 23, 2002. 

This event was observed by the IZMIRAN group  on 21 May, 2002. The CR counts reduction observed here occured slowly but with sharp recovery phases in all the five stations. The IZMIRAN group from a network of Neutron monitors assigned a value of 1.9 \% to the magnitude of this event \citep{be:2018}. The result for the magnitude of this event for Inuvik NM station is obtained to be 2.4 \% using the manual approach technique expressed as \citep[e.g.][]{har:2010, lee:2013, oki_:2020}  

\begin{equation} \label{eqn}
FD_ {mag} (\%) = \frac {CR_{max}-CR_{onset}}{CR_{mean}}\times 100 ,
\end{equation}\\
where $FD_{mag} (\%)$ is the FD magnitude, $CR_{max}$ is the CR-count at the end of the main phase, $CR_{onset}$ is the CR-count at the onset of the main phase and $CR_{mean}$ represents the mean value of the CR data for a predefined period.\\

Figure 2 is a graph illustrating the time profile of non-simultaneous  FD event recorded by  six NM stations: Jungfraujoch, Climax, Irkutsk, Kerguelen, Oulu and Magadan. This event was observed on  12 to  13 February, 2000. 

 Two groups of NMs observed the same FD event at different times. Irkutsk, Magadan and Oulu recorded the event on February 12 while Kerguelen, Jungfraujoch and Climax detected it on  February 13. The two events on February 12 and 13 appear to show rapid CR intensity reduction and slow recovery. We estimated the magnitude of FD at IRKT station through equation 1 to be 3.7\%. The IZMIRAN group  identified this event on February 11 with FD value of 4.2\%.\\

\begin{figure} 
\begin{minipage}[c]{0.44\linewidth}
\includegraphics[width=\linewidth]{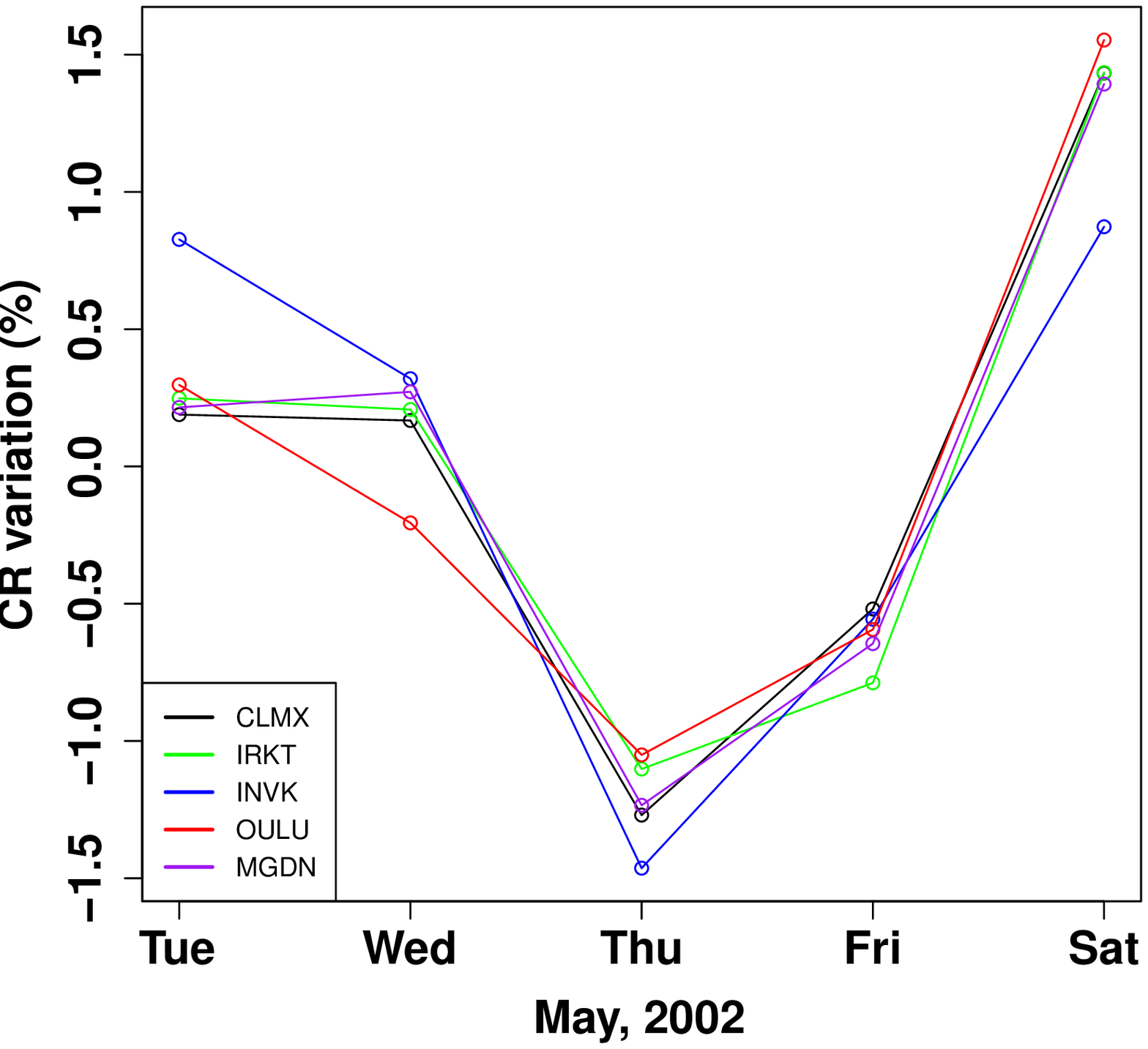}
\caption{\textbf{CR flux profiles of simultaneous FD event of  23 May, 2002}}
\end{minipage}
\hfill
\begin{minipage}[c]{0.44\linewidth}
\includegraphics[width=\linewidth]{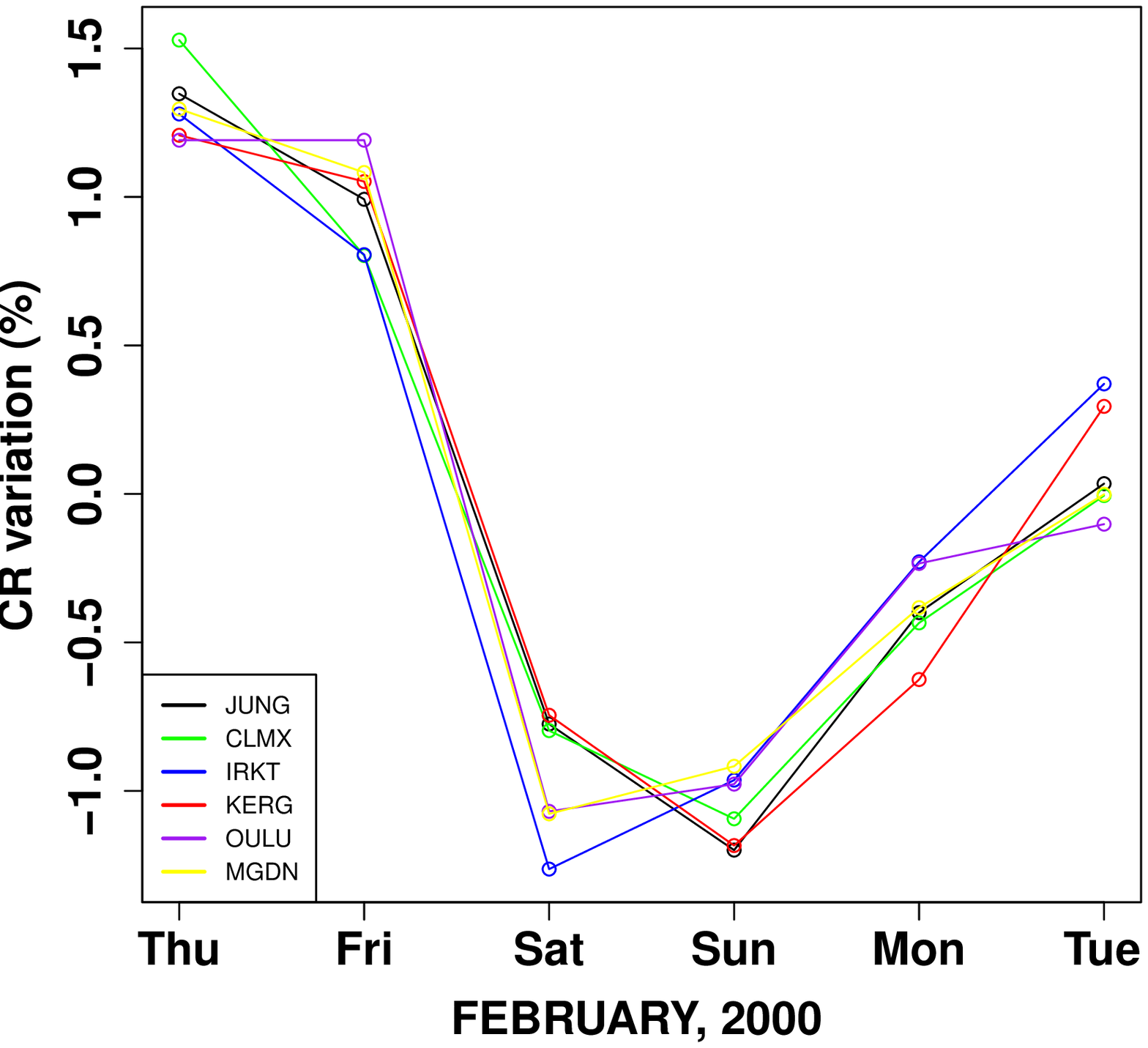}
\caption{\textbf{CR flux profiles of Non-simultaneous FD event of   12 - 13 February, 2000}}
\end{minipage}
\end{figure}
\onecolumn
\begin{table}[ht]
\caption{ Forbush Decreases from INVK (FD1\%), MGDN (FD2\%) and OULU (FD3\%)}
\label{Table 2}
\scriptsize
\centering
\begin{tabular}{rlrlrlr}
  \hline
 S/N& Date1 & FD1\% & Date2 & FD2\% & Date3 & FD3\% \\ 
  \hline
1 & 1998-08-27 & -1.63 & 1998-08-27 & -1.53 & 1998-08-27 & -2.62 \\ 
  2 & 1998-09-25 & -0.57 & 1998-09-25 & -0.16 & 1998-09-25 & -0.75 \\ 
  3 & 1999-01-24 & -1.07 & 1999-02-18 & -0.29 & 1999-01-24 & -0.67 \\ 
  4 & 1999-02-18 & -2.95 & 1999-08-23 & -0.11 & 1999-02-18 & -2.23 \\ 
  5 & 1999-08-22 & -1.07 & 1999-10-12 & -0.31 & 1999-08-22 & -0.53 \\ 
  6 & 1999-08-25 & -0.03 & 1999-10-16 & -1.43 & 1999-10-17 & -0.90 \\ 
  7 & 1999-10-17 & -0.82 & 1999-10-22 & -1.82 & 1999-10-22 & -0.72 \\ 
  8 & 1999-10-22 & -0.95 & 1999-10-25 & -1.36 & 1999-10-24 & -0.85 \\ 
  9 & 1999-10-25 & -1.04 & 1999-10-28 & -0.41 & 1999-11-12 & -0.18 \\ 
  10 & 1999-11-14 & -0.56 & 1999-10-31 & -0.71 & 1999-11-17 & -0.64 \\ 
  11 & 1999-11-17 & -0.96 & 1999-11-13 & -0.93 & 1999-11-20 & -1.06 \\ 
  12 & 1999-11-20 & -1.60 & 1999-11-18 & -0.99 & 1999-11-22 & -0.87 \\ 
  13 & 1999-12-03 & -0.51 & 1999-11-23 & -1.82 & 1999-12-03 & -1.53 \\ 
  14 & 1999-12-13 & -4.85 & 1999-12-01 & -1.27 & 1999-12-13 & -5.51 \\ 
  15 & 1999-12-22 & -0.87 & 1999-12-03 & -1.21 & 1999-12-27 & -2.70 \\ 
  16 & 1999-12-27 & -2.25 & 1999-12-13 & -4.89 & 1999-12-31 & -0.75 \\ 
  17 & 2000-01-01 & -0.26 & 1999-12-21 & -0.40 & 2000-01-02 & -0.70 \\ 
  18 & 2000-01-04 & -0.39 & 1999-12-27 & -2.23 & 2000-01-04 & -0.79 \\ 
  19 & 2000-01-07 & -0.37 & 2000-01-02 & -0.31 & 2000-01-06 & -0.75 \\ 
  20 & 2000-01-20 & -0.23 & 2000-01-05 & -0.34 & 2000-01-24 & -0.82 \\ 
  21 & 2000-01-24 & -1.74 & 2000-01-07 & -0.64 & 2000-01-29 & -0.05 \\ 
  22 & 2000-01-28 & -0.37 & 2000-01-13 & -0.27 & 2000-02-07 & -0.55 \\ 
  23 & 2000-01-30 & -0.33 & 2000-01-24 & -0.64 & 2000-02-12 & -3.76 \\ 
  24 & 2000-02-07 & -0.22 & 2000-02-01 & -0.27 & 2000-02-21 & -2.13 \\ 
  25 & 2000-02-09 & -0.05 & 2000-02-08 & -0.64 & 2000-03-01 & -2.73 \\ 
  26 & 2000-02-13 & -3.90 & 2000-02-12 & -4.00 & 2000-03-04 & -1.83 \\ 
  27 & 2000-02-21 & -2.72 & 2000-02-22 & -2.21 & 2000-03-06 & -1.59 \\ 
  28 & 2000-03-01 & -3.61 & 2000-03-01 & -2.94 & 2000-03-09 & -1.81 \\ 
  29 & 2000-03-09 & -1.62 & 2000-03-13 & -2.78 & 2000-03-13 & -2.62 \\ 
  30 & 2000-03-13 & -2.49 & 2000-03-20 & -1.99 & 2000-03-19 & -2.26 \\ 
  31 & 2000-03-16 & -2.33 & 2000-03-25 & -4.03 & 2000-03-24 & -3.59 \\ 
  32 & 2000-03-20 & -1.90 & 2000-03-28 & -3.43 & 2000-03-30 & -3.15 \\ 
  33 & 2000-03-24 & -3.86 & 2000-03-30 & -3.70 & 2000-04-05 & -2.99 \\ 
  34 & 2000-03-28 & -3.16 & 2000-04-01 & -2.95 & 2000-04-07 & -4.36 \\ 
  35 & 2000-03-30 & -3.48 & 2000-04-04 & -3.34 & 2000-04-14 & -0.79 \\ 
  36 & 2000-04-02 & -2.42 & 2000-04-07 & -5.19 & 2000-04-17 & -1.76 \\ 
  37 & 2000-04-04 & -2.63 & 2000-04-17 & -2.30 & 2000-04-19 & -1.98 \\ 
  38 & 2000-04-07 & -4.38 & 2000-04-20 & -2.81 & 2000-04-22 & -1.79 \\ 
  39 & 2000-04-14 & -1.31 & 2000-04-24 & -3.37 & 2000-04-24 & -2.30 \\ 
  40 & 2000-04-17 & -1.87 & 2000-05-03 & -4.47 & 2000-05-03 & -3.59 \\ 
  41 & 2000-04-20 & -2.56 & 2000-05-09 & -5.42 & 2000-05-08 & -4.21 \\ 
  42 & 2000-04-24 & -2.58 & 2000-05-12 & -3.28 & 2000-05-16 & -3.71 \\ 
  43 & 2000-04-29 & -1.65 & 2000-05-15 & -4.77 & 2000-05-24 & -7.87 \\ 
  44 & 2000-05-03 & -3.96 & 2000-05-17 & -4.22 & 2000-05-30 & -4.38 \\ 
  45 & 2000-05-08 & -4.98 & 2000-05-24 & -8.77 & 2000-06-09 & -9.69 \\ 
  46 & 2000-05-15 & -4.03 & 2000-05-30 & -4.58 & 2000-06-20 & -5.94 \\ 
  47 & 2000-05-24 & -8.22 & 2000-06-09 & -10.35 & 2000-06-24 & -6.01 \\ 
  48 & 2000-05-30 & -4.09 & 2000-06-20 & -6.56 & 2000-06-26 & -6.19 \\ 
  49 & 2000-06-02 & -3.87 & 2000-06-24 & -6.45 & 2000-07-02 & -3.32 \\ 
  50 & 2000-06-09 & -10.29 & 2000-06-26 & -6.14 & 2000-07-05 & -3.34 \\ 
  51 & 2000-06-20 & -6.37 & 2000-07-04 & -4.12 & 2000-07-08 & -3.20 \\ 
  52 & 2000-06-24 & -6.05 & 2000-07-06 & -4.08 & 2000-07-11 & -5.86 \\ 
  53 & 2000-07-02 & -3.90 & 2000-07-11 & -6.45 & 2000-07-13 & -8.33 \\ 
  54 & 2000-07-06 & -3.90 & 2000-07-16 & -15.77 & 2000-07-16 & -15.58 \\ 
  55 & 2000-07-11 & -5.81 & 2000-07-20 & -11.15 & 2000-07-20 & -10.88 \\ 
  56 & 2000-07-13 & -8.49 & 2000-07-29 & -8.74 & 2000-07-22 & -10.21 \\ 
  57 & 2000-07-16 & -15.27 & 2000-08-06 & -8.30 & 2000-07-29 & -7.92 \\ 
  58 & 2000-07-20 & -11.07 & 2000-08-12 & -9.89 & 2000-08-06 & -7.97 \\ 
  59 & 2000-07-29 & -8.66 & 2000-08-15 & -8.28 & 2000-08-12 & -8.86 \\ 
  60 & 2000-08-06 & -8.19 & 2000-08-25 & -4.68 & 2000-08-26 & -4.46 \\
  61 & 2000-08-12 & -9.83 & 2000-08-29 & -4.89 & 2000-08-31 & -3.93 \\ 
  62 & 2000-08-25 & -4.21 & 2000-09-03 & -4.39 & 2000-09-03 & -4.18 \\ 
  63 & 2000-08-29 & -4.32 & 2000-09-07 & -5.08 & 2000-09-09 & -5.61 \\ 
  64 & 2000-09-02 & -5.46 & 2000-09-12 & -5.38 & 2000-09-15 & -5.47 \\ 
  65 & 2000-09-08 & -6.72 & 2000-09-15 & -5.88 & 2000-09-18 & -9.59 \\ 
  66 & 2000-09-12 & -6.28 & 2000-09-18 & -9.72 & 2000-09-26 & -3.89 \\ 
  67 & 2000-09-20 & -7.67 & 2000-09-26 & -3.97 & 2000-09-29 & -4.01 \\ 
  68 & 2000-09-25 & -4.66 & 2000-09-29 & -3.97 & 2000-10-01 & -4.16 \\ 
  69 & 2000-09-29 & -4.49 & 2000-10-01 & -4.06 & 2000-10-05 & -4.33 \\ 
  70 & 2000-10-01 & -4.66 & 2000-10-05 & -4.61 & 2000-10-07 & -4.58 \\ 
  71 & 2000-10-04 & -5.01 & 2000-10-08 & -4.27 & 2000-10-14 & -3.61 \\ 
  72 & 2000-10-08 & -4.74 & 2000-10-13 & -3.20 & 2000-10-20 & -1.81 \\ 
  73 & 2000-10-14 & -4.90 & 2000-10-18 & -2.57 & 2000-10-22 & -1.71 \\ 
  74 & 2000-10-20 & -2.89 & 2000-10-20 & -2.43 & 2000-10-29 & -6.09 \\ 
  75 & 2000-10-29 & -7.90 & 2000-10-26 & -1.78 & 2000-11-04 & -4.48 \\
\hline
\end{tabular}
\end{table}

\begin{table}[ht]
\label*{table 2}
\scriptsize
\centering
\begin{tabular}{rlrlrlr}
  \hline
  S/N& Date1 & FD1\% & Date2 & FD2\% & Date3 & FD3\% \\ 
  \hline 
 
  76 & 2000-11-01 & -6.28 & 2000-10-29 & -6.78 & 2000-11-07 & -7.79 \\ 
  77 & 2000-11-03 & -5.77 & 2000-11-01 & -5.51 & 2000-11-11 & -6.21 \\ 
  78 & 2000-11-07 & -9.00 & 2000-11-04 & -5.12 & 2000-11-13 & -4.90 \\ 
  79 & 2000-11-11 & -7.39 & 2000-11-07 & -8.07 & 2000-11-16 & -4.92 \\ 
  80 & 2000-11-14 & -6.84 & 2000-11-11 & -6.43 & 2000-11-24 & -4.40 \\ 
  81 & 2000-11-23 & -6.02 & 2000-11-14 & -5.32 & 2000-11-29 & -9.77 \\ 
  82 & 2000-11-29 & -11.30 & 2000-11-29 & -10.72 & 2000-12-06 & -6.19 \\ 
  83 & 2000-12-14 & -4.31 & 2000-12-03 & -6.78 & 2000-12-11 & -4.20 \\ 
  84 & 2000-12-19 & -4.99 & 2000-12-23 & -4.23 & 2000-12-15 & -3.17 \\ 
  85 & 2000-12-23 & -5.53 & 2000-12-26 & -4.55 & 2000-12-19 & -3.29 \\ 
  86 & 2000-12-27 & -5.83 & 2000-12-30 & -3.13 & 2000-12-23 & -4.58 \\ 
  87 & 2000-12-30 & -4.52 & 2001-01-02 & -3.15 & 2000-12-25 & -4.60 \\ 
  88 & 2001-01-03 & -4.48 & 2001-01-05 & -2.91 & 2000-12-27 & -4.90 \\ 
  89 & 2001-01-05 & -4.14 & 2001-01-09 & -3.09 & 2000-12-31 & -3.51 \\ 
  90 & 2001-01-09 & -4.94 & 2001-01-15 & -2.81 & 2001-01-03 & -3.78 \\ 
  91 & 2001-01-18 & -5.11 & 2001-01-18 & -3.32 & 2001-01-10 & -3.69 \\ 
  92 & 2001-01-24 & -5.92 & 2001-01-25 & -4.20 & 2001-01-14 & -3.37 \\ 
  93 & 2001-01-29 & -4.94 & 2001-02-01 & -3.10 & 2001-01-19 & -3.14 \\ 
  94 & 2001-01-31 & -4.94 & 2001-02-10 & -1.31 & 2001-01-25 & -4.67 \\ 
  95 & 2001-02-06 & -2.25 & 2001-02-14 & -1.59 & 2001-01-31 & -2.92 \\ 
  96 & 2001-02-11 & -2.25 & 2001-02-20 & -1.05 & 2001-02-07 & -1.84 \\ 
  97 & 2001-02-14 & -2.77 & 2001-03-05 & -0.10 & 2001-02-11 & -1.24 \\ 
  98 & 2001-02-20 & -1.63 & 2001-03-20 & -0.23 & 2001-02-14 & -2.03 \\ 
  99 & 2001-02-27 & -0.75 & 2001-03-28 & -2.81 & 2001-02-20 & -1.22 \\ 
  100 & 2001-03-04 & -1.34 & 2001-04-01 & -3.69 & 2001-03-04 & -1.46 \\ 
  101 & 2001-03-20 & -0.78 & 2001-04-05 & -4.80 & 2001-03-21 & -0.57 \\ 
  102 & 2001-04-01 & -6.19 & 2001-04-09 & -6.49 & 2001-03-28 & -2.73 \\ 
  103 & 2001-04-05 & -6.98 & 2001-04-12 & -12.63 & 2001-04-01 & -5.22 \\ 
  104 & 2001-04-08 & -8.13 & 2001-04-16 & -5.87 & 2001-04-05 & -5.25 \\ 
  105 & 2001-04-12 & -13.93 & 2001-04-19 & -4.76 & 2001-04-09 & -6.43 \\ 
  106 & 2001-04-16 & -6.02 & 2001-04-22 & -3.26 & 2001-04-12 & -12.79 \\ 
  107 & 2001-04-19 & -4.73 & 2001-04-25 & -1.63 & 2001-04-16 & -5.57 \\ 
  108 & 2001-04-22 & -3.42 & 2001-04-29 & -6.30 & 2001-04-19 & -4.60 \\ 
  109 & 2001-04-26 & -2.24 & 2001-05-04 & -3.76 & 2001-04-22 & -3.05 \\ 
  110 & 2001-04-30 & -6.95 & 2001-05-08 & -2.44 & 2001-04-26 & -1.95 \\ 
  111 & 2001-05-04 & -3.70 & 2001-05-15 & -2.62 & 2001-04-29 & -6.92 \\ 
  112 & 2001-05-08 & -2.18 & 2001-05-20 & -1.78 & 2001-05-08 & -1.63 \\ 
  113 & 2001-05-14 & -2.28 & 2001-05-25 & -3.66 & 2001-05-12 & -1.93 \\ 
  114 & 2001-05-25 & -3.68 & 2001-05-28 & -5.24 & 2001-05-16 & -1.42 \\ 
  115 & 2001-05-28 & -4.80 & 2001-06-02 & -2.84 & 2001-05-20 & -0.84 \\ 
  116 & 2001-06-03 & -2.81 & 2001-06-05 & -1.87 & 2001-05-25 & -2.95 \\ 
  117 & 2001-06-10 & -2.63 & 2001-06-07 & -2.09 & 2001-05-28 & -5.30 \\ 
  118 & 2001-06-12 & -2.46 & 2001-06-10 & -2.40 & 2001-06-03 & -2.11 \\ 
  119 & 2001-06-18 & -2.42 & 2001-06-12 & -2.41 & 2001-06-09 & -1.88 \\ 
  120 & 2001-06-20 & -2.92 & 2001-06-18 & -1.91 & 2001-06-12 & -1.83 \\ 
  121 & 2001-06-26 & -2.22 & 2001-06-20 & -2.36 & 2001-06-20 & -1.93 \\ 
  122 & 2001-06-30 & -1.73 & 2001-06-26 & -2.06 & 2001-06-26 & -1.26 \\ 
  123 & 2001-07-04 & -2.81 & 2001-06-30 & -1.67 & 2001-06-29 & -1.17 \\ 
  124 & 2001-07-08 & -2.44 & 2001-07-05 & -2.63 & 2001-07-05 & -2.16 \\ 
  125 & 2001-07-17 & -0.95 & 2001-07-08 & -1.95 & 2001-07-09 & -1.58 \\ 
  126 & 2001-07-20 & -1.27 & 2001-07-11 & -1.50 & 2001-07-17 & -0.27 \\ 
  127 & 2001-07-23 & -2.07 & 2001-07-16 & -0.34 & 2001-07-19 & -0.53 \\ 
  128 & 2001-07-25 & -2.27 & 2001-07-20 & -1.30 & 2001-07-21 & -0.32 \\ 
  129 & 2001-07-30 & -2.02 & 2001-07-24 & -2.01 & 2001-07-26 & -1.63 \\ 
  130 & 2001-08-03 & -3.00 & 2001-07-27 & -2.07 & 2001-07-30 & -1.51 \\ 
  131 & 2001-08-07 & -2.58 & 2001-07-30 & -1.64 & 2001-08-03 & -2.43 \\ 
  132 & 2001-08-13 & -2.35 & 2001-08-02 & -2.64 & 2001-08-06 & -2.50 \\ 
  133 & 2001-08-18 & -4.03 & 2001-08-06 & -2.51 & 2001-08-14 & -1.56 \\ 
  134 & 2001-08-23 & -3.42 & 2001-08-09 & -1.88 & 2001-08-18 & -3.76 \\ 
  135 & 2001-08-28 & -7.25 & 2001-08-13 & -2.27 & 2001-08-24 & -3.05 \\ 
  136 & 2001-09-07 & -2.98 & 2001-08-18 & -4.69 & 2001-08-29 & -7.02 \\ 
  137 & 2001-09-14 & -1.54 & 2001-08-23 & -3.62 & 2001-09-07 & -2.72 \\ 
  138 & 2001-09-19 & -1.24 & 2001-08-29 & -7.49 & 2001-09-15 & -1.69 \\ 
  139 & 2001-09-26 & -7.37 & 2001-09-07 & -2.99 & 2001-09-19 & -1.46 \\ 
  140 & 2001-09-30 & -7.71 & 2001-09-14 & -1.85 & 2001-09-22 & -0.92 \\ 
  141 & 2001-10-09 & -4.31 & 2001-09-16 & -1.75 & 2001-09-26 & -7.92 \\ 
  142 & 2001-10-12 & -7.09 & 2001-09-19 & -1.38 & 2001-10-01 & -7.81 \\ 
  143 & 2001-10-22 & -6.19 & 2001-09-21 & -1.47 & 2001-10-09 & -4.38 \\ 
  144 & 2001-10-28 & -5.04 & 2001-09-26 & -7.08 & 2001-10-12 & -5.76 \\ 
  145 & 2001-11-06 & -6.28 & 2001-09-30 & -7.55 & 2001-10-22 & -5.22 \\ 
  146 & 2001-11-14 & -2.02 & 2001-10-02 & -8.61 & 2001-10-28 & -4.68 \\ 
  147 & 2001-11-22 & -2.83 & 2001-10-09 & -4.82 & 2001-11-07 & -6.75 \\ 
  148 & 2001-11-25 & -7.14 & 2001-10-12 & -6.49 & 2001-11-15 & -2.16 \\ 
  149 & 2001-12-01 & -0.89 & 2001-10-22 & -5.12 & 2001-11-22 & -3.07 \\ 
  150 & 2001-12-06 & -3.22 & 2001-10-28 & -4.79 & 2001-11-25 & -8.43 \\
  151 & 2001-12-17 & -2.72 & 2001-11-03 & -0.79 & 2001-12-01 & -1.46 \\ 
  152 & 2001-12-21 & -1.51 & 2001-11-07 & -6.09 & 2001-12-06 & -3.51 \\
\hline
\end{tabular}
\end{table}

\begin{table}[ht]
\label*{table 2}
\scriptsize
\centering
\begin{tabular}{rlrlrlr}
  \hline
  S/N& Date1 & FD1\% & Date2 & FD2\% & Date3 & FD3\% \\ 
  \hline
 
  153 & 2001-12-29 & -2.30 & 2001-11-14 & -1.73 & 2001-12-17 & -3.39 \\ 
  154 & 2002-01-03 & -7.26 & 2001-11-22 & -2.95 & 2001-12-22 & -1.02 \\ 
  155 & 2002-01-12 & -6.47 & 2001-11-25 & -8.72 & 2001-12-24 & -0.67 \\ 
  156 & 2002-01-19 & -3.23 & 2001-12-06 & -3.53 & 2001-12-27 & -1.29 \\ 
  157 & 2002-01-21 & -3.39 & 2001-12-17 & -3.00 & 2001-12-29 & -2.25 \\ 
  158 & 2002-01-30 & -5.41 & 2001-12-22 & -0.99 & 2002-01-03 & -7.29 \\ 
  159 & 2002-02-01 & -5.32 & 2001-12-25 & -0.39 & 2002-01-11 & -6.63 \\ 
  160 & 2002-02-12 & -1.23 & 2002-01-01 & -6.55 & 2002-01-19 & -3.12 \\ 
  161 & 2002-02-14 & -1.18 & 2002-01-03 & -6.67 & 2002-01-21 & -3.24 \\ 
  162 & 2002-02-18 & -1.32 & 2002-01-12 & -6.06 & 2002-01-24 & -2.99 \\ 
  163 & 2002-02-23 & -2.35 & 2002-01-21 & -3.41 & 2002-01-30 & -5.27 \\ 
  164 & 2002-02-26 & -2.02 & 2002-01-23 & -3.34 & 2002-02-01 & -5.09 \\ 
  165 & 2002-02-28 & -1.96 & 2002-01-29 & -4.71 & 2002-02-11 & -0.77 \\ 
  166 & 2002-03-02 & -1.83 & 2002-02-01 & -4.98 & 2002-02-15 & -0.42 \\ 
  167 & 2002-03-06 & -1.35 & 2002-02-12 & -0.73 & 2002-02-19 & -0.75 \\ 
  168 & 2002-03-12 & -1.68 & 2002-02-19 & -0.84 & 2002-02-23 & -1.89 \\ 
  169 & 2002-03-16 & -1.54 & 2002-02-23 & -1.91 & 2002-03-02 & -2.26 \\ 
  170 & 2002-03-25 & -6.39 & 2002-02-25 & -1.78 & 2002-03-05 & -1.83 \\ 
  171 & 2002-03-30 & -4.15 & 2002-02-28 & -1.85 & 2002-03-12 & -1.76 \\ 
  172 & 2002-04-06 & -1.90 & 2002-03-02 & -1.35 & 2002-03-16 & -1.63 \\ 
  173 & 2002-04-12 & -2.16 & 2002-03-05 & -0.64 & 2002-03-22 & -5.14 \\ 
  174 & 2002-04-15 & -2.11 & 2002-03-07 & -0.21 & 2002-03-25 & -6.29 \\ 
  175 & 2002-04-18 & -5.52 & 2002-03-12 & -1.63 & 2002-03-30 & -3.51 \\ 
  176 & 2002-04-20 & -4.98 & 2002-03-16 & -1.45 & 2002-04-05 & -1.74 \\ 
  177 & 2002-04-24 & -4.90 & 2002-03-23 & -5.40 & 2002-04-12 & -2.72 \\ 
  178 & 2002-05-08 & -1.12 & 2002-03-25 & -5.42 & 2002-04-15 & -1.58 \\ 
  179 & 2002-05-13 & -2.33 & 2002-03-30 & -3.86 & 2002-04-18 & -4.06 \\ 
  180 & 2002-05-16 & -2.81 & 2002-04-12 & -2.90 & 2002-04-20 & -5.19 \\ 
  181 & 2002-05-20 & -4.31 & 2002-04-15 & -2.30 & 2002-04-24 & -5.10 \\ 
  182 & 2002-05-23 & -6.16 & 2002-04-18 & -5.39 & 2002-05-04 & -0.69 \\ 
  183 & 2002-05-27 & -4.24 & 2002-04-20 & -5.74 & 2002-05-08 & -1.11 \\ 
  184 & 2002-06-03 & -2.95 & 2002-04-22 & -4.98 & 2002-05-13 & -2.89 \\ 
  185 & 2002-06-11 & -3.14 & 2002-04-24 & -6.54 & 2002-05-15 & -3.10 \\ 
  186 & 2002-06-19 & -2.92 & 2002-04-30 & -2.69 & 2002-05-20 & -4.45 \\ 
  187 & 2002-06-24 & -1.87 & 2002-05-03 & -1.22 & 2002-05-23 & -5.19 \\ 
  188 & 2002-06-29 & -0.89 & 2002-05-12 & -3.46 & 2002-05-28 & -4.11 \\ 
  189 & 2002-07-03 & -1.69 & 2002-05-15 & -3.74 & 2002-06-04 & -2.20 \\ 
  190 & 2002-07-09 & -3.05 & 2002-05-21 & -5.12 & 2002-06-07 & -2.26 \\ 
  191 & 2002-07-11 & -2.77 & 2002-05-23 & -6.70 & 2002-06-12 & -2.84 \\ 
  192 & 2002-07-18 & -4.32 & 2002-05-27 & -4.43 & 2002-06-16 & -1.86 \\ 
  193 & 2002-07-20 & -6.16 & 2002-06-03 & -2.64 & 2002-06-19 & -2.94 \\ 
  194 & 2002-07-25 & -5.32 & 2002-06-07 & -2.42 & 2002-06-24 & -1.32 \\ 
  195 & 2002-07-30 & -8.12 & 2002-06-11 & -3.57 & 2002-06-29 & -0.58 \\ 
  196 & 2002-08-02 & -9.05 & 2002-06-19 & -2.73 & 2002-07-03 & -1.79 \\ 
  197 & 2002-08-20 & -7.03 & 2002-06-24 & -1.67 & 2002-07-11 & -3.10 \\ 
  198 & 2002-08-23 & -6.67 & 2002-06-28 & -0.93 & 2002-07-18 & -4.38 \\ 
  199 & 2002-08-29 & -7.40 & 2002-07-03 & -1.52 & 2002-07-20 & -6.13 \\ 
  200 & 2002-09-04 & -5.78 & 2002-07-09 & -2.97 & 2002-07-23 & -5.20 \\ 
  201 & 2002-09-08 & -6.16 & 2002-07-11 & -3.06 & 2002-08-02 & -9.12 \\ 
  202 & 2002-09-10 & -5.32 & 2002-07-18 & -3.91 & 2002-08-07 & -6.01 \\ 
  203 & 2002-09-19 & -4.09 & 2002-07-20 & -5.61 & 2002-08-09 & -5.45 \\ 
  204 & 2002-09-24 & -5.01 & 2002-07-23 & -4.87 & 2002-08-20 & -6.71 \\ 
  205 & 2002-09-28 & -4.59 & 2002-07-30 & -8.19 & 2002-08-23 & -6.60 \\ 
  206 & 2002-10-01 & -4.24 & 2002-08-02 & -8.63 & 2002-08-28 & -7.65 \\ 
  207 & 2002-10-03 & -4.37 & 2002-08-20 & -6.91 & 2002-09-04 & -5.61 \\ 
  208 & 2002-10-13 & -1.77 & 2002-08-23 & -6.06 & 2002-09-08 & -6.01 \\ 
  209 & 2002-10-21 & -6.30 & 2002-08-29 & -7.17 & 2002-09-11 & -5.22 \\ 
  210 & 2002-10-25 & -5.11 & 2002-09-02 & -4.91 & 2002-09-19 & -4.06 \\ 
  211 & 2002-10-27 & -5.02 & 2002-09-08 & -5.69 & 2002-09-24 & -5.19 \\ 
  212 & 2002-10-30 & -3.82 & 2002-09-11 & -4.81 & 2002-09-28 & -4.48 \\ 
  213 & 2002-11-03 & -4.54 & 2002-09-23 & -4.63 & 2002-10-01 & -4.62 \\ 
  214 & 2002-11-05 & -5.44 & 2002-09-28 & -3.76 & 2002-10-03 & -5.17 \\ 
  215 & 2002-11-12 & -6.28 & 2002-10-01 & -3.46 & 2002-10-09 & -2.35 \\ 
  216 & 2002-11-18 & -7.87 & 2002-10-03 & -4.06 & 2002-10-13 & -2.13 \\ 
  217 & 2002-11-27 & -4.82 & 2002-10-08 & -1.41 & 2002-10-21 & -6.03 \\ 
  218 & 2002-12-01 & -3.54 & 2002-10-13 & -1.40 & 2002-10-25 & -5.37 \\ 
  219 & 2002-12-08 & -4.68 & 2002-10-21 & -5.29 & 2002-11-03 & -5.07 \\ 
  220 & 2002-12-15 & -5.16 & 2002-11-03 & -5.03 & 2002-11-05 & -6.21 \\ 
  221 & 2002-12-18 & -5.24 & 2002-11-05 & -5.81 & 2002-11-12 & -7.13 \\ 
  222 & 2002-12-20 & -5.41 & 2002-11-12 & -7.15 & 2002-11-18 & -9.02 \\ 
  223 & 2002-12-23 & -6.64 & 2002-11-18 & -8.08 & 2002-11-25 & -4.43 \\ 
  224 & 2002-12-25 & -6.22 & 2002-11-25 & -3.70 & 2002-11-27 & -5.52 \\ 
  225 &  &  & 2002-11-28 & -4.44 & 2002-12-01 & -4.41 \\ 
  226 &  &  & 2002-12-08 & -4.40 & 2002-12-06 & -3.84 \\ 
  227 &  &  & 2002-12-15 & -4.53 & 2002-12-08 & -5.02 \\ 
  228 &  &  & 2002-12-20 & -5.44 & 2002-12-15 & -5.61 \\ 
  229 &  &  & 2002-12-23 & -6.40 & 2002-12-20 & -6.70 \\ 
  230 &  &  &  &  & 2002-12-23 & -7.47 \\ 
   \hline
\end{tabular}
\end{table}

\twocolumn
\begin{figure}[!b]
  \begin{center}
    \includegraphics[width=3.5in]{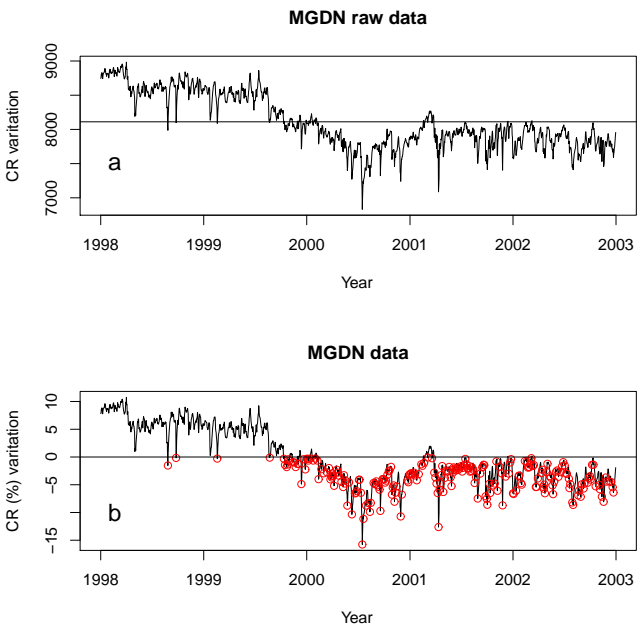}
  \end{center}

  \caption{\small Raw (panel a) and normalized (panel b) CR data from Magadan  NM station.The static mean
cosmic ray data  is shown with the horizontal line.
The 
dips are signatures of FDs. 
      }
  \label{MGDN}
\end{figure}

\begin{figure}[!b]
  \begin{center}
    \includegraphics[width=3.5in]{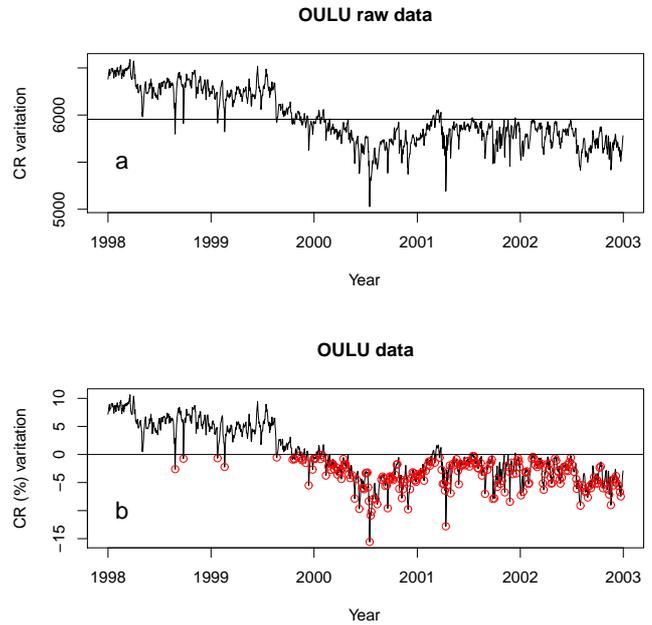}
  \end{center}

  \caption{\small Raw (panel a) and normalized (panel b) CR data from Oulu  NM station. The static mean
cosmic ray data  is shown with the horizontal line. The 
dips are indicators of  FDs.
  }
  \label{OULU}
\end{figure}

\begin{figure}[!b]
  \begin{center}
    \includegraphics[width=3.5in]{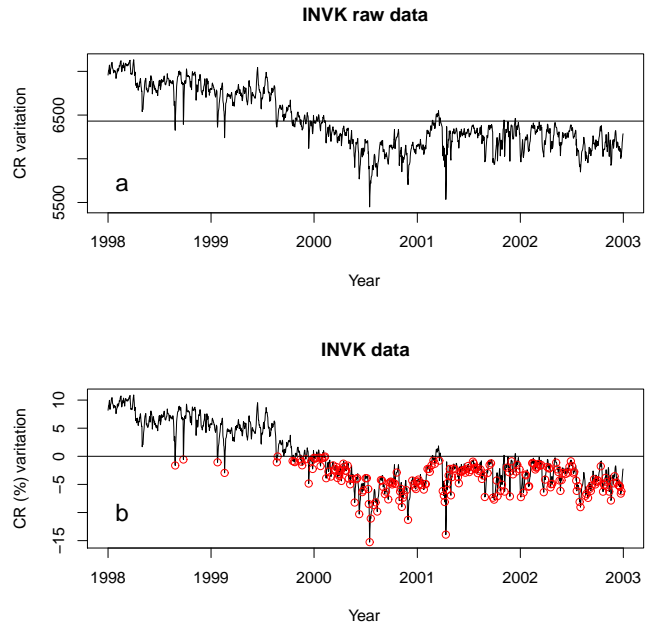}
  \end{center}

  \caption{\small Raw (panel a) and normalized (panel b) CR data from Inuvik  NM station. The static mean
cosmic ray data  is shown with the horizontal line. The 
dips show FDs. 
      }
  \label{INVK}
\end{figure}

\section{Analysis}

\subsection{Techniques for Detection and Timing of CR Transient Events}\label{Techniques}
Precise timing and sensitive detection of transient events such as X-ray photons, CMEs, GLEs and FDs remain one of the challenges in all aspects of astrophysics. In addition to the conventional manual/visual event identification technique, astrophysicists have developed and implemented several detection algorithms \citep[see][for example]{Yashiro:2008,lamy:2019} including the recently introduced Bayesian blocks type algorithm \citep{Scargle:1998,Way:2011,Scargle:2013,Worple:2015}. While each of the software tools developed to detect astrophysical transient events are designed for specific data, the Bayesian block or changepoint algorithm is a non-parametric event detection method, initially developed for unit events as photons. It has been speculated that it may be applied to composite events such as FDs as well. Changepoint detection aims to identify points in time series data at which the statistical properties of a sequence of observations change. This algorithm is implemented in the changepoint package in R programming software. The method is usually implemented for the change in mean and/or variance \citep{Killick:2014}.

The result of the Bayesian blocks algorithm allows the researcher to visualize the locations of significant changes in the amplitude, and widths of the pulse structure within a burst. The detected pulse features may be used as a starting point for parametric or quantitative test. Nevertheless, a survey of the literature indicates that the Bayesian blocks method is not widely used to detect CR transient events. This could be due to the presence of unknown and the observed periodicities resulting from diurnal variations as well as inherent cycles and recurrence-tendencies in CR data \citep{bart:1935}. The sporadic nature of composite CR events like FDs may also limit the usefulness of Bayesian Blocks analysis. To extract such periodic or quasi-periodic signals in CR time-intensity variations demands lots of time averaging as is the case in the traditional Fourier and power spectrum analyses. \citet{Scargle:1998} noted that changepoint method of analysis may not be appropriate for such data.  

Here, we briefly review the two popular methods of FD event identification.

\subsection{Manual Event identification Method}
This method involves plotting the time-sorted CR data and visually searching for points of intensity reductions in the data. The complete time profile of a FD is usually defined by four parts comprising of the event onset, main phase, point of maximal reduction and the recovery phase as schematically illustrated in Figure 1 of \citet{oh:08}. The next step, after determining the event time profile, involves calculation of the event magnitude and timing. As indicated before, equation \ref{eqn} is generally used to estimate the amplitude of reduction. Unfortunately, event timing is not accounted for in equation \ref{eqn}, implying that the researcher would have to search for the time of occurrence of the event, either with respect to the onset or time of maximal reduction. Although a series of FDs may happen within a short interval as clearly illustrated in Figure 1 of \citet{ok:2020}, each of the steps outlined here must be repeated for each event.

In spite of the laborious and time consuming nature of the exercise, large volumes of FD catalogues including those of \citet{lo:1990}, \cite{tin:1991}, \citet{ca:93}, \citet{ca:96}, \citet{sv:09}, \citet{laken:2011}, \citet{sven:2016} and the IZMIRAN have been created in the past by the manual approach. Although many space-weather investigations have been conducted based on these catalogues, the inherent flaws in the selection methodology has been the focus of some recent articles \citep[see][for instance]{ok:2019c,ok:2020,ok:2021b}.

\subsection{Algorithm-based identification Method}
As indicated in the introductory section, there has been a little progress here. The approach of \citet{rami:2013} and the recent effort of \citet{Light:2020} are semi-automated selection method somewhat similar to those of the IZMIRAN team. While an algorithm is implemented at some initial stage to mark the presence of FDs in CR data \citep[see Figure 1 of][for example]{Light:2020}, event magnitude estimation and timing is manually carried out. All the semi-automated method adopt a case study approach. Here, irrespective of the volume of the data under investigation, the magnitude and timing of the FDs are calculated separately.

Besides the versions of our Fourier-R-based FD location algorithm which has be used in various applications such as \citet{ok:2019}, \citet{ok:2019a}, \citet{ok:2019b}, \citet{ok:2020} and \citet{ok:2021}, we know of no other publications that have fully automated Forbush event identification.   



\subsubsection {R-based FD Location  Algorithm}

The current FD-location code is adapted from a version newly developed and implemented by \citet{ok:2021b}. The code is written in an R package \citep[see][]{R:2014}. R is a open source code developed by Robert Gentleman and Ross Ihaka of the statistical department, University of Auckland and maintained by collaborators across the globe. While the software can be freely downloaded from https://www.r-project.org/, interested readers willing to use the software may register at r-help@r-project.org in other to work with several experts that freely assist R-users.

The algorithm is capable of detecting weak and significant instantaneous increases as well as decreases in time series data. Where the interest is on intensity reductions as in the case of FDs, the subroutines that tracks event peaks are disabled, allowing only for the detection of troughs/pits/minima in the data. The code is extremely sensitive as it is designed to identify a very small time intensity changes throughout the data. It can also handle both small and extremely large volume time series data. 

Besides allowing us to mark and visualize the changepoints (as is usually attempted with the Bayesian blocks approach, the GSM, IDL protocol of \citet{rami:2013}, and the recent FD detection code developed by \citet{Light:2020}) associated with intensity reductions (or increases when focusing on GLEs), our algorithm simultaneously calculates the magnitude and the  event time of occurrence and writes the result to a file. It is interesting to note that both the marking/detection, timing, calculation of low/high event amplitudes, and cataloguing processes are fully automated. Our method has other advantages over other transient detection approaches. For example, while Bayesian blocks method neither uses explicit smoothing nor pre-defined binning (omni-scale), for example, one of the beautiful features of the current FD identification code is that it can be applied to both unsmoothed and smoothed data (multi-scale). This allows us to make detailed comparisons between the results of FD event simultaneity tests based on daily and hourly count rates in the present article.

Despite the merits of the current code, it has some limitations. One of the pitfalls is the inability of the algorithm to account for the contributions of CR anisotropy in the time-intensity variations of both the daily and hourly CR data. \citet{ba:74} observed that small amplitude FDs are easily obscured by diurnal CR anisotropies and may not be detected without some sort of data transformation. We hope to address this flaw in a future work.

\section{Results and Discussion}
\subsection{FDs Detected by R-FD Location Algorithm}
The FD event magnitude and time of occurrence calculated by the location algorithm are presented in Table 2 for the three stations. Our  code picked 229, 230 and 224 FDs from Magadan, Oulu and Inuvik NMs, respectively, whose amplitudes range from 0.01\% to 15.77\%. The columns are explained thus: S/N stand for serial number, Date1, Date2 and Date3, respectively represent FD dates for INVK, MGDN and OULU, while FD1(\%), FD2(\%) and FD3(\%) denote FD amplitude/size for INVK, MGDN and OULU respectively. In the three stations, CR data points (i.e. number of days for which each station has data from 1998 to 2002) are  as follows: MGDN (1813/1825), OULU (1823/1825) and INVK (1802/1825). We infer that the minor differeces between the FDs selected by the present code from the three stations might be due to the marginal differences in the data length of MGDN, OULU and INVK NM detectors. Other factors like geomagnetic cut off rigidity, altitude, rotation of the Earth with respect to the acceptance cone of the detectors, instrumental variations could also be responsible  \citep[e.g.][]{oki:20}. The algorithm selected slightly more sample of FDs from Oulu than the rest stations consistent with the data lengths.

For the same period (1998-2002), the IZMIRAN group selected a total  of 662 FD events from their hourly data. The FD catalogue obtained with the present code from Oulu station is compared with the FD catalogue 
 of the IZMIRAN group. The R-FD algorithm selected 34.7\% of the IZMIRAN group FD catalogue \citep{be:2018}. In the manual method, small magnitude FDs are difficult to pick in a given CR count data as against the results obtained with our algorithm. The difficulty has been linked to the eclipsing nature of diurnal anisotropies \citep{ba:75}. Detecting such large number of small intensity depressions comparable to large events in the present data attest to the efficiency of the location code employed in this work.\\ 

\subsection{Simultaneous FD Sample} 

To  identify FDs that are  simultaneous  at the  three NMs, we employed a  simple R-FD coincident algorithm code. FD coincident algorithm is a simple program written to select the FDs commonly observed by the neutron monitor detectors. The code scans two or more column data with respect to a specified key search parameter. The simultaneous FDs observed at INVK and MGDN, INVK and OULU and MGDN and OULU are respectively 138, 138 and 125. 

For the three stations, the test of  simultaneity   was further carried out.  
These simultaneous FD events were selected  with respect to the amplitude of FDs at MGDN station. The input data to the coincident algorithm are all the
FDs at each of the stations: MGDN = 229, OULU = 230, INVK = 224 (see Table \ref{Table 2}). FD event with amplitude = 0.11\% at MGDN was chosen as key search index. The result   show that a total of 99 FDs made up  of small and large FD event sizes occured at the same day at the three stations (see Table \ref{Table 3}). The columns are organized as follows: S/N, represent  serial number, FD1(\%), FD2(\%) and FD3(\%)  stand for MGDN, OULU and INVK FDs respectively. 
The strongest FD event occurred on July 16, 2000. This was observed in the three stations. Our FD location technique assigned 15.77\%, 15.58\% and 15.27\% to MGDN, OULU and INUVK stations respectively. MGDN NM recorded the largest FD magnitude in this event, while INVK trailed behind OULU station. The second large ranking event was picked by our code on Apri 12, 2001. For the three stations, the values are respectively MGDN 12.63\%, OULU 12.79\% and INVK 13.93\%. This time, the FD amplitude in INVK station  rank higer than those in the other two stations. The third ranking FD event was selected by the code on July 20, 2000. The magnitudes corresponding  to the three stations (MGDN, OULU and INVK) are 11.15\%, 10.88\% and 11.07\%. MGDN has the highest FD amplitude in this event with OULU station recording the lowest of the three. The event of September 25, 1998 is the smallest of the simultaneous FDs. The magnitude of the FD recorded at OULU NM (0.75\%) is higher than those at MGDN (0.16\%) and INVK (0.57\%). 

The few striking events highlighted could suffice for the reader to follow through. We do not intend to discuss all the simultaneous FD events in this work. That will be taken up in future work. The differences in FD amplitudes observed at different NMs revealed in this result might be  related to differences in
Forbush event manifestations and their location dependent effects \citep[e.g.][]{be:08}. 

The event of July 16, 2000 observed by the coincident algorithm is very outstanding. \citet{oh:08} and \citet{kris:08} respectively assigned FD magnitude of 10.33\% and 13\% for this event. The IZMIRAN group from their technique of network of NMs obtained a value of 13.60\% for this event at the onset time on July 15, 2000 \citep{be:2015}.\\  
For the simultaneous FD on April 12, 2001, \citep{oh:08} calculated a FD magnitude of 11.82\% while \citep{kris:08} obtained a magnitude of 12\%. The event was recorded by the IZMIRAN team on April 11, 2001 with a value of 14.00\%. Third ranking event of July 20, 2000 was not in the catalogue of both \citep{oh:08} and \citep{kris:08}. It was detected by the IZMIRAN team on onset date of July 19, 2000 with a value of 4.3\%. This value is higher than the size identified by the present code. The least  ranking FD on September 25, 1998 was seen  by the IZMIRAN group on the main phase onset date of September 24, 1998 with an amplitude of 8.7\%. This result is larger than the size estimated with the current algorithm.  In general, the closeness in magnitude of the FD results from isolated NM stations  and a network of NM detectors demonstrate the efficiency of the current code developed and deployed in this work. 

Employing the hourly averaged data of the Oulu NMs from 1998 to 2002, \citep{oh:08}  selected 49 FD events with  a threshold of $>$ 3.5\% GCR intensity decrease. By comparing Oulu NM station data with the data from Inuvik and Magadan NM stations, they showed that 37 out of 49 FDs are detected by the three separate stations simultaneously in universal time. \citet{kris:08}  using a baseline of $\leq$ 5\%  selected 22 large FDs within a six year period from 2000 and 2005 using Climax data. The dates of these events were compared with those  with FD days found in  Oulu and Moscow NMs.  Our algorithm on the other hand  selected 69 simultaneous FDs of comparable magnitude within a five year period from 1998 to 2002 from OULU NMs. This result underscore the merit of algorithm-selected FDs over  manual FD detection approach.
The result obtained here indicates that the code used for FD detection in this work is  efficient. The result will be of immense value to FD CR scientists in the study of solar-terrestrial connections. 

A number of past articles that use the manual FD selection method often found that large FDs are simultaneously detected at the same UT, while small amplitude FDs are non-simultaneous events \citep[e.g.][]{oh:08, oh:09, le:2015, Melkumyan:2018}. But the current code selected both small and large FDs in the simultaneous FD catalogue \citep{oki_:2020}. \\


\subsection{Result Validation}

\begin{itemize}
	\item Bayesian Blocks Algorithm
 
\end{itemize}

In order to confirm the discussions or conclusions based on the current R-FD search algorithm, result validation is of topmost priority. There are a number of reasons for this. First, the use of daily (smoothed) rather than hourly data constitutes an apparent drawback on the results of the analysis. Hourly counts are generally employed in several important studies such as CME/ICME-FD related analysis \citep[e.g.][]{ca:00,Jordan:2011} and investigation of CR diurnal vectors \citep[see][and the references therein]{Borie:2016} where timing accuracy is a desideratum. Second, the 11-year solar cycle contribution, which is evident in Figures 3, 4 and 5 is not accounted for. \citet{har:2011} indicate the need to remove effect of solar cycle variations from CR data before FD event selection. Third, CR anisotropy, known to exert significant influence on the number, magnitude, and timing of Forbush events \citep{Pomerantz:1971, be:08,oki_:2020,ok:2021} was not accounted for in the current work. 

Implementation of a Bayesian blocks type algorithm on the hourly CR data may allow us verify whether the outcome of FD event simultaneity presented here are real CR time-intensity variation effects as previously suggested by \citet{oh:08} and \citet{ok:2011} or the result of a binning/averaging artefact arising from statistical fluctuation in the data. However, since the Bayesian blocks analysis are generally not reported in the literature analyzing CR data, the unsmoothed data from one of the stations (OULU) is first tested for suitability. The raw (hourly) data spanning the same period (1998-2002) of the daily was used. The hourly data are individual events measured as a time series rather then the binned (daily) counts. 

The hourly data was first cleaned by interpolating zero (missing) values within the observed values. Using the changepoint package in the R software, a total number of 32779 changepoints were detected. This is very large compared to the number of data points (43825) in the raw data. In order to visualize the locations of the changepoints, we projected the locations of the changepoints on the raw data (result not shown). The changepoints seem to run from the first data points to the end of the data.

 The hourly counts were further fitted to a gamma distribution in order to test whether Bayesian blocks analysis is suitable for analysis of binned CR counts. Over-plotted histograms of the observed and theoretical distributions were generated and a goodness of fit test conducted using the Chi-squared statistics. The Chi-squared (${\chi}^2$), the degree of freedom (DF), and the associated p-value are respectively 158.57, 143, and 0.1741. Table D.4 of \citet{guj:2004} indicates that the DF is not statistically significant at any level. The non-significant p-value indicated here also reflects the extremely large changepoints (32779) reported for OULU station between 1998 and 2002. Inspection of existing works equally confirms that the number of FDs that happened within this five year period is far too small compared with the current too many changepoints. For example, \citet{oh:08}, \citet{laken:2011} and IZMIRAN respectively reported a total of 49, 16 and 704 FDs within the same period. In the light of the small number of FDs detected by different researchers using empirical data, it is improbable that a total changepoints of 32779 are indicators of FDs. Practically, GLEs and several other short-term intensity fluctuations, including the periodicities and recurrences in the CR data \citep{bart:1935}, might contaminate or bias the result of changepoint or Bayesian block type analysis when applied to raw/unprocessed CR data. 

In the light of these multiple signal superposition inherent in raw CR data (see Section \ref{Techniques}), it is not surprising that the changepoint search algorithm is not suitable for analysis of CR time series data. The result of Bayesian algorithm may be confounded by the competing superposed signals.  While FD event detection, timing and magnitude estimation are the interests of the current work, for example, the Bayesian blocks algorithm which tracks the locations of instantaneous increases or decreases in count rate \citep{Worple:2015} may not discriminate between FDs and other divers CR intensity changes such as GLEs, and anomalous enhancement of CRs intensity variations \citep{Dorman:2018}. In the next section, attempt will be made to validate the presented result using a modified version of the current R-FD detection algorithm.

\begin{itemize}
	\item Application of the Modified R-FD Location Algorithm to Hourly Data
 
\end{itemize}

Records of CR data yield functions of time such as hourly, bi-hourly and daily \citep{shea:2000} and are usually transformed into data series of equal time intervals such as hourly, daily, monthly, annually and so on. Several CR properties have also been investigated on these different time scales. Hourly values are usually converted to daily data using a 24-hour running mean \citep[see][for example]{Ruffolo:2016}. There are cases where analysis based on short-time scales are the same with those over longer periods, whereas they are situations the outcome is significantly different. In particular, a number of investigators \citep[e.g][]{lo:71,Dumbovic:2011,Ruffolo:2016,be:2018} have alluded to the differences in CR phenomena for hourly and daily resolutions. The statistical accuracy of hourly count and daily count rates are respectively 0.1 and 0.02\% \citep{shea:2000,Ruffolo:2016} respectively. \citet{be:2018} and the references therein also highlighted the disadvantages as well as advantages of using daily averages CR data.

In view of the above-mentioned differences between hourly and daily averages, hourly data from the three stations are employed to validate our result of FD event simultaneity. In order to achieve this, the current version of our FD identification code was adjusted to accept hourly NM records as the input signal. The new code was then used to calculate the FD event time as well as magnitude. A total number of 4055(42396), 4032(43269), and 4144(43772) FDs were selected from INVK, MGDN, and OULU stations respectively. The number enclosed in the brackets represent the hours each station made observation for the year 1998-2002. While a larger baseline ($\mathrm{CR} (\%)\leq -0.01$) was employed for the daily data, a smaller baseline ($\mathrm{CR} (\%)\leq -3$) was used for the hourly values \citep[see][]{ok:2019c,ok:2021c,ok:2021d}.

\begin{figure}[h!]
 \centering
  \includegraphics[width=1\textwidth]{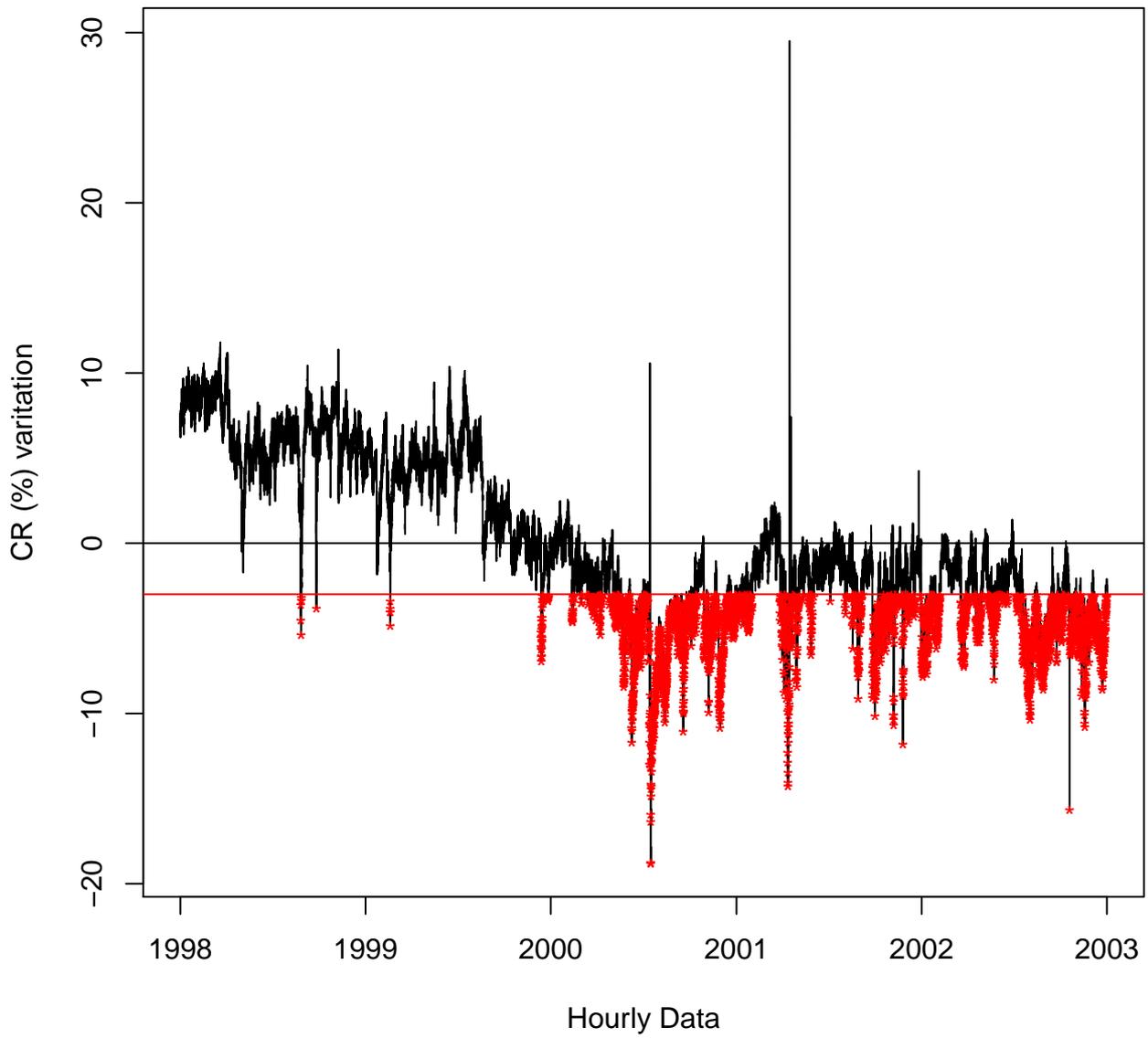}
 \caption{\textbf{Normalized CR count rate from OULU station. The red stars are indications of FDs}}
	\label{ALL_OULU}
 \end{figure}

Figure \ref{ALL_OULU} presents the hourly data for OULU station. The red horizontal line and the red stars respectively represent the normalization baseline and the FDs. The upward spikes are markers of GLEs that occurred within the period \citep[see Table 1 of][]{Usoskin:2011}. For the sake of brevity, similar results for INVK and MGDN are not presented.

Figure \ref{ALL_OULU} and panel b of Figure \ref{OULU} allows for comparison of some of the differences between hourly and daily CR count rates. It is evident from the two diagrams that shorter time duration events like GLEs are adversely affected by the binning or averaging effects. This could explain the absence of spikes in panel b of Figure \ref{OULU}. While the averaging procedure may not have the same weight on some FDs, a careful comparison of the two diagrams shows that it, nevertheless, registers significant impact on the number, magnitude, and ultimately, on the timing of FDs. The contribution from the solar cycle oscillation is also noticeable in the two diagrams. 
 
Using the coincident algorithm, we searched for simultaneous FDs among the 4055, 4032, and 4144 FDs identified at INVK, MGDN, and OULU stations. A total of 261 simultaneous FDs were identified. The result is presented in Table \ref{Table 4}. A comparison of Tables \ref{Table 4} and \ref{Table 3}(simultaneous events based on daily resolutions) shows that some of the differences or similarities between the daily and hourly count rates reflected in Figure \ref{ALL_OULU} and panel b of Figure \ref{OULU} are also apparent in the two Tables. 

The result presented in Table \ref{Table 4} seems to strengthen the outcome of FD event simultaneity test conducted using the daily means. Nevertheless, the differences in the number of simultaneous FDs presented in the two Tables are significant, suggesting that averaging procedure may play a key role on the global event simultaneity. The number of simultaneous FDs selected using the hourly data is greater than those detected with the daily averages by about a factor of 3. If to consider the event timing in the two Tables, one would also notice some significant differences. While simultaneous FDs appear in Table \ref{Table 3} for 1998 and 1999, no event happened  simultaneously in the two years in Table \ref{Table 4}. Although several simultaneous FDs are recorded in the years of high solar activities (Table \ref{Table 4}), a careful comparison of the events in 2000-2002 indicates significant timing differences between the two data resolutions.  

Besides the binning effects, we also tested whether the vertical geomagnetic cutoff rigidity (which is different for different detectors) also influences FD event simultaneity. One of the quick checks is to compare the magnitude of the simultaneous FDs at the three stations. For the daily data, the average magnitude of the simultaneous event at INVK, MGDN and OULU are respectively 4.85, 4.70 and 4.67\% respectively. It is 5.61, 5.40 and 5.35\% for INVK, MGDN and OULU stations. It is interesting to note that INVK which has the lowest cutoff rigidity consistently measures higher magnitude. This may be a pointer that geomagnetic cutoff rigidity impacts on FD event simultaneity. But the indicated connection between magnitude and rigidity contrasts a little with OULU station, which has a lower rigidity than MGDN, but measures a somewhat lower magnitude in comparison. Higher altitude of MGDN and several other factors affecting NM sensitivity, as noted by \citet{ok:2020b}, may also play a role. 

Before making definitive statements on the differences and similarities between FD simultaneity tests presented using the daily and hourly averages, there is a need to account for the impact of CR diurnal anisotropy \citep[see][]{ok:2021d} and the solar cycle variations on the CR data from the separate NMs. Using daily CR data from INVK, OULU and MGDN, \citet{ok:2020b} demonstrated the implications of solar cycle variations on the amplitudes and number of FD for Solar cycle 23. A comparison of Figure \ref{OULU} of this work and Figure 4 of \citet{ok:2021b} clearly shows that solar cycle variations could have some pronounced effects on the manifestations of Forbush events. Using daily count rates from Climax station for the period of 1953 to 2006 (Solar Cycle 19-23), the method of removing solar cycle effect and other hidden periodicities in CR data has also been presented \citep[see Figure 1 of][]{ok:2021c}.

\section{Summary}
Detection and precise timing of transient events like photons, CMEs/ICMEs, FDs and GLEs frequently raise issues that remain on the cutting edge of research in astrophysics. This is particularly true in solar-terrestrial studies where FDs are used to time the arrival of solar event sources on the Earth's magnetosphere. While significant progress has been recorded with regard to automation of CME, and photon detection, for example, manual method has been predominantly used to identify FDs and other CR transient events. Whereas the manual method may be applicable for case studies of FD events as well as for analysis of CR data series that span a few years \citep[see][]{ba:73,ba:74,oh:08}, the current work shows that it may be grossly inadequate in the face of a large volume of data.  

Detection of small or weak FD signals is one of the major flaws of the traditional manual selection technique. Accurate detection and timing of weak signals are some of the factors that motivated the implementation of Bayesian blocks algorithm in astrophysics \citep{Scargle:2013}. Nevertheless, the method may be unsuitable for some data distribution like ours where lots of signal superimposition is involved.

 While the GSM of the IZMIRAN group has a detection accuracy of 0.3-0.5\% for hourly count rates \citep[e.g][]{be:2018,be:2018b}, Table \ref{Table 2} shows that the current algorithm registers a detection accuracy of $\leq$ 0.01\% for daily averaged CR data. Using the unsmoothed data, we further demonstrated the detection efficiency of our code. For the first time, FD lists selected using the same methodical approach from smoothed and unsmoothed CR data are used to investigate the global simultaneity of Forbush events. As the simultaneous FDs represented in Tables \ref{Table 3} and \ref{Table 4} only differ in temporal resolutions, their comparison is not confounded with other factors like detector type, data type, methodology, and so on \citep[see][for details on factors that make unbiased comparison or uniting of different FD lists impossible]{Abunin:2013}.  

Though certain challenges such as accounting for the impacts of CR anisotropy and solar cycle oscillation on the physical properties of FDs remain, the global FD event simultaneity result presented using the hourly NM count rates showcases the current version of our R-FD location algorithm as an extremely sensitive statistical technique. While lending credence to the outcome of FD event simultaneity test conducted with the daily means, the significant differences between the two catalogues are apparent cautionary signs to several researchers \citep[e.g][]{kr:08,la:2009} that conduct FD simultaneity/consistent test using daily count rates. 

Compared with the large number of FDs detected at the three stations, the number of events that are simultaneously measured at the three stations are quite few. This is a pointer to the significantly different flux variations seen at the different locations. Although \citet{Ruffolo:2016} claim that short-term CR flux changes can be measured using a single NM, the simultaneity test presented here shows that shor-term time-intensity variations may differ significantly at different points on the Earth.

\section{Conclusion}
A new automated FD location algorithm has been used to develop FD catalogue from the daily-averaged  CR raw data  at Magadan, Oulu and Inuvik NM stations spanning the period 1998 to 2002. The coincident code detected a total of 99 FDs including small and large amplitude events from the algorithm-selected FDs sample that are recorded at the  three  stations. The large number of FD events selected in five years compared to the few FDs detected over  many years in some previous works may point to the advantage of automated computer program. The closeness in amplitude of the FD list from isolated NM stations  and a network of NM detectors demonstrate the efficiency of the current program. Although the coincident code deployed in this work clearly demonstrates simultaneous FD event detection at the three stations, there is a need to test the validity of the presented results with hourly CR data. 

The outcome of FD event selection and simultaneity test using hourly CR values are significantly different from those based on the daily CR data. The number of FDs selected using hourly and daily averages are in the ratio (hourly/daily) of about 18:1 for each of the three stations. Though the number of FDs identified from the hourly data are comparatively larger than those selected from the daily means, it is interesting to note that the number of simultaneous FDs detected from the daily averages ($\approx$ 43\% of the total event) are relatively greater than those identified from the hourly data ($\approx$ 6\% of the total) at the three NMs. 

The observed difference may be a pointer to a binning/averaging artefact in the daily data. We equally speculate that CR diurnal anisotropy, expected to be averaged out or greatly reduced in daily CR means may play a key role in the observed number difference between simultanous FDs. The phase shift arising from enhanced CR diurnal anisotropy in the unsmoothed data may be the reason for the relative fewer number of simultaneous FDs.  
\onecolumn

\begin{table}[ht]
\caption{ Simultaneous FD events from MGDN (FD1\%), OULU (FD2\%) and INVK (FD3\%)}
\label{Table 3}
\centering
\begin{tabular}{rlrrr}
  \hline
 S/N & Date & FD1\% & FD2\% & FD3\% \\ 
  \hline
1 & 1998-08-27 & -1.53 & -2.62 & -1.63 \\ 
  2 & 1998-09-25 & -0.16 & -0.75 & -0.57 \\ 
  3 & 1999-02-18 & -0.29 & -2.23 & -2.95 \\ 
  4 & 1999-10-22 & -1.82 & -0.72 & -0.95 \\ 
  5 & 1999-12-03 & -1.21 & -1.53 & -0.51 \\ 
  6 & 1999-12-13 & -4.89 & -5.51 & -4.85 \\ 
  7 & 1999-12-27 & -2.23 & -2.70 & -2.25 \\ 
  8 & 2000-01-24 & -0.64 & -0.82 & -1.74 \\ 
  9 & 2000-03-01 & -2.94 & -2.73 & -3.61 \\ 
  10 & 2000-03-13 & -2.78 & -2.62 & -2.49 \\ 
  11 & 2000-03-30 & -3.70 & -3.15 & -3.48 \\ 
  12 & 2000-04-07 & -5.19 & -4.36 & -4.38 \\ 
  13 & 2000-04-17 & -2.30 & -1.76 & -1.87 \\ 
  14 & 2000-04-24 & -3.37 & -2.30 & -2.58 \\ 
  15 & 2000-05-03 & -4.47 & -3.59 & -3.96 \\ 
  16 & 2000-05-24 & -8.77 & -7.87 & -8.22 \\ 
  17 & 2000-05-30 & -4.58 & -4.38 & -4.09 \\ 
  18 & 2000-06-09 & -10.35 & -9.69 & -10.29 \\ 
  19 & 2000-06-20 & -6.56 & -5.94 & -6.37 \\ 
  20 & 2000-06-24 & -6.45 & -6.01 & -6.05 \\ 
  21 & 2000-07-11 & -6.45 & -5.86 & -5.81 \\ 
  22 & 2000-07-16 & -15.77 & -15.58 & -15.27 \\ 
  23 & 2000-07-20 & -11.15 & -10.88 & -11.07 \\ 
  24 & 2000-07-29 & -8.74 & -7.92 & -8.66 \\ 
  25 & 2000-08-06 & -8.30 & -7.97 & -8.19 \\ 
  26 & 2000-08-12 & -9.89 & -8.86 & -9.83 \\ 
  27 & 2000-09-29 & -3.97 & -4.01 & -4.49 \\ 
  28 & 2000-10-01 & -4.06 & -4.16 & -4.66 \\ 
  29 & 2000-10-20 & -2.43 & -1.81 & -2.89 \\ 
 \hline
\end{tabular}
\end{table}

\begin{table}[ht]
\label*{table 3}
\centering
\begin{tabular}{rlrrr}
 S/N & Date & FD1\% & FD2\% & FD3\% \\ 
  \hline 
  30 & 2000-10-29 & -6.78 & -6.09 & -7.90 \\
  31 & 2000-11-07 & -8.07 & -7.79 & -9.00 \\ 
  32 & 2000-11-11 & -6.43 & -6.21 & -7.39 \\ 
  33 & 2000-11-29 & -10.72 & -9.77 & -11.30 \\ 
  34 & 2000-12-23 & -4.23 & -4.58 & -5.53 \\ 
  35 & 2001-02-14 & -1.59 & -2.03 & -2.77 \\ 
  36 & 2001-02-20 & -1.05 & -1.22 & -1.63 \\ 
  37 & 2001-04-01 & -3.69 & -5.22 & -6.19 \\ 
  38 & 2001-04-05 & -4.80 & -5.25 & -6.98 \\ 
  39 & 2001-04-12 & -12.63 & -12.79 & -13.93 \\ 
  40 & 2001-04-16 & -5.87 & -5.57 & -6.02 \\ 
  41 & 2001-04-19 & -4.76 & -4.60 & -4.73 \\ 
  42 & 2001-04-22 & -3.26 & -3.05 & -3.42 \\ 
  43 & 2001-05-08 & -2.44 & -1.63 & -2.18 \\ 
  44 & 2001-05-25 & -3.66 & -2.95 & -3.68 \\ 
  45 & 2001-05-28 & -5.24 & -5.30 & -4.80 \\ 
  46 & 2001-06-12 & -2.41 & -1.83 & -2.46 \\ 
  47 & 2001-06-20 & -2.36 & -1.93 & -2.92 \\ 
  48 & 2001-06-26 & -2.06 & -1.26 & -2.22 \\ 
  49 & 2001-07-30 & -1.64 & -1.51 & -2.02 \\ 
  50 & 2001-08-18 & -4.69 & -3.76 & -4.03 \\ 
  51 & 2001-09-07 & -2.99 & -2.72 & -2.98 \\ 
  52 & 2001-09-19 & -1.38 & -1.46 & -1.24 \\ 
  53 & 2001-09-26 & -7.08 & -7.92 & -7.37 \\ 
  54 & 2001-10-09 & -4.82 & -4.38 & -4.31 \\ 
  55 & 2001-10-12 & -6.49 & -5.76 & -7.09 \\ 
  56 & 2001-10-22 & -5.12 & -5.22 & -6.19 \\ 
  57 & 2001-10-28 & -4.79 & -4.68 & -5.04 \\ 
  58 & 2001-11-22 & -2.95 & -3.07 & -2.83 \\ 
  59 & 2001-11-25 & -8.72 & -8.43 & -7.14 \\  
 \hline
\end{tabular}
\end{table}

\begin{table}[ht]
\label*{table 3}
\centering
\begin{tabular}{rlrrr}
  \hline
 S/N & Date & FD1\% & FD2\% & FD3\% \\ 
  \hline
  60 & 2001-12-06 & -3.53 & -3.51 & -3.22 \\
  61 & 2001-12-17 & -3.00 & -3.39 & -2.72 \\ 
  62 & 2002-01-03 & -6.67 & -7.29 & -7.26 \\ 
  63 & 2002-01-21 & -3.41 & -3.24 & -3.39 \\ 
  64 & 2002-02-01 & -4.98 & -5.09 & -5.32 \\ 
  65 & 2002-02-23 & -1.91 & -1.89 & -2.35 \\ 
  66 & 2002-03-02 & -1.35 & -2.26 & -1.83 \\ 
  67 & 2002-03-12 & -1.63 & -1.76 & -1.68 \\ 
  68 & 2002-03-16 & -1.45 & -1.63 & -1.54 \\ 
  69 & 2002-03-25 & -5.42 & -6.29 & -6.39 \\ 
  70 & 2002-03-30 & -3.86 & -3.51 & -4.15 \\ 
  71 & 2002-04-12 & -2.90 & -2.72 & -2.16 \\ 
  72 & 2002-04-15 & -2.30 & -1.58 & -2.11 \\ 
  73 & 2002-04-18 & -5.39 & -4.06 & -5.52 \\ 
  74 & 2002-04-20 & -5.74 & -5.19 & -4.98 \\ 
  75 & 2002-04-24 & -6.54 & -5.10 & -4.90 \\ 
  76 & 2002-05-23 & -6.70 & -5.19 & -6.16 \\ 
  77 & 2002-06-19 & -2.73 & -2.94 & -2.92 \\ 
  78 & 2002-06-24 & -1.67 & -1.32 & -1.87 \\ 
  79 & 2002-07-03 & -1.52 & -1.79 & -1.69 \\ 
  80 & 2002-07-11 & -3.06 & -3.10 & -2.77 \\ 
  81 & 2002-07-18 & -3.91 & -4.38 & -4.32 \\ 
  82 & 2002-07-20 & -5.61 & -6.13 & -6.16 \\ 
  83 & 2002-08-02 & -8.63 & -9.12 & -9.05 \\ 
  84 & 2002-08-20 & -6.91 & -6.71 & -7.03 \\ 
  85 & 2002-08-23 & -6.06 & -6.60 & -6.67 \\ 
  86 & 2002-09-08 & -5.69 & -6.01 & -6.16 \\ 
  87 & 2002-09-28 & -3.76 & -4.48 & -4.59 \\ 
  88 & 2002-10-01 & -3.46 & -4.62 & -4.24 \\ 
  89 & 2002-10-03 & -4.06 & -5.17 & -4.37 \\ 
 
 \hline
\end{tabular}
\end{table}

\begin{table}[ht]
\label*{table 3}
\centering
\begin{tabular}{rlrrr}
  \hline
 S/N & Date & FD1\% & FD2\% & FD3\% \\ 
  \hline
  90 & 2002-10-13 & -1.40 & -2.13 & -1.77 \\ 
  91 & 2002-10-21 & -5.29 & -6.03 & -6.30 \\ 
  92 & 2002-11-03 & -5.03 & -5.07 & -4.54 \\ 
  93 & 2002-11-05 & -5.81 & -6.21 & -5.44 \\ 
  94 & 2002-11-12 & -7.15 & -7.13 & -6.28 \\ 
  95 & 2002-11-18 & -8.08 & -9.02 & -7.87 \\ 
  96 & 2002-12-08 & -4.40 & -5.02 & -4.68 \\ 
  97 & 2002-12-15 & -4.53 & -5.61 & -5.16 \\ 
  98 & 2002-12-20 & -5.44 & -6.70 & -5.41 \\ 
  99 & 2002-12-23 & -6.40 & -7.47 & -6.64 \\ 
   \hline
\end{tabular}
\end{table}

\onecolumn

\begin{table}[ht]
\caption{ Simultaneous FD events from INVK (FD1\%), OULU (FD2\%) and MGDN(FD3\%) selected from hourly data}
\label{Table 4}
\centering
\begin{tabular}{rlrrrr}
  \hline
 S/N& Date & Hr & FD1\% & FD2\% & FD3\% \\ 
  \hline
1 & 2000-02-12 &   2 & -4.05 & -4.25 & -3.79 \\ 
  2 & 2000-03-25 &   8 & -4.46 & -3.74 & -4.88 \\ 
  3 & 2000-03-30 &   0 & -3.17 & -3.44 & -3.42 \\ 
  4 & 2000-05-02 &   9 & -3.40 & -3.10 & -5.25 \\ 
  5 & 2000-05-08 &   1 & -4.98 & -4.82 & -5.64 \\ 
  6 & 2000-05-12 &   0 & -3.06 & -3.02 & -3.82 \\ 
  7 & 2000-05-15 &   3 & -3.49 & -4.45 & -4.95 \\ 
  8 & 2000-05-16 &   1 & -4.04 & -4.40 & -4.43 \\ 
  9 & 2000-05-20 &   3 & -4.80 & -3.39 & -4.99 \\ 
  10 & 2000-05-24 &   1 & -8.27 & -8.19 & -8.44 \\ 
  11 & 2000-05-24 &   1 & -7.89 & -8.18 & -8.71 \\ 
  12 & 2000-05-27 &   6 & -5.65 & -6.29 & -6.43 \\ 
  13 & 2000-05-27 &   1 & -5.14 & -5.20 & -6.21 \\ 
  14 & 2000-06-01 &   9 & -3.88 & -3.61 & -4.32 \\ 
  15 & 2000-06-02 &   2 & -3.82 & -3.69 & -4.21 \\ 
  16 & 2000-06-02 &   3 & -4.42 & -3.44 & -4.64 \\ 
  17 & 2000-06-03 &   0 & -3.65 & -3.74 & -3.68 \\ 
  18 & 2000-06-13 &   2 & -7.29 & -7.64 & -7.42 \\ 
  19 & 2000-06-15 &   3 & -6.49 & -5.79 & -6.86 \\ 
  20 & 2000-06-16 &   8 & -5.20 & -5.49 & -6.24 \\ 
  21 & 2000-06-17 &   0 & -5.02 & -4.48 & -5.55 \\ 
  22 & 2000-06-20 &   2 & -6.01 & -6.83 & -6.22 \\ 
  23 & 2000-06-21 &   7 & -6.45 & -6.21 & -6.17 \\ 
  24 & 2000-06-23 &   0 & -4.46 & -6.66 & -5.25 \\ 
  25 & 2000-06-23 &   2 & -5.22 & -3.39 & -5.36 \\ 
  26 & 2000-06-24 &   2 & -6.56 & -6.13 & -6.55 \\ 
  27 & 2000-06-25 &   6 & -5.89 & -5.07 & -5.97 \\ 
  28 & 2000-06-28 &   7 & -5.37 & -5.25 & -6.01 \\ 
  29 & 2000-07-02 &   2 & -3.93 & -3.74 & -4.09 \\ 
  30 & 2000-07-02 &   6 & -4.27 & -3.32 & -4.31 \\ 
  31 & 2000-07-04 &   2 & -3.29 & -3.99 & -3.66 \\ 
  32 & 2000-07-04 &   8 & -4.18 & -3.10 & -4.26 \\ 
  33 & 2000-07-05 &   6 & -4.10 & -3.41 & -4.38 \\ 
  34 & 2000-07-07 &   3 & -3.99 & -3.52 & -4.10 \\ 
  35 & 2000-07-07 &   6 & -4.08 & -3.10 & -4.00 \\ 
  36 & 2000-07-15 &   1 & -10.32 & -10.71 & -11.45 \\ 
  37 & 2000-07-19 &   3 & -10.51 & -9.65 & -11.00 \\ 
  38 & 2000-07-20 &   7 & -11.27 & -11.45 & -11.76 \\ 
  39 & 2000-07-23 &   2 & -10.01 & -10.86 & -9.88 \\ 
  40 & 2000-07-24 &   2 & -10.02 & -8.23 & -10.31 \\ 
  41 & 2000-07-29 &   6 & -8.56 & -8.02 & -9.46 \\ 
  42 & 2000-07-30 &   8 & -8.61 & -7.10 & -8.41 \\ 
  43 & 2000-07-31 &   3 & -5.87 & -6.35 & -6.03 \\ 
  44 & 2000-08-01 &   0 & -6.32 & -5.88 & -6.44 \\ 
  45 & 2000-08-02 &   3 & -6.56 & -6.13 & -5.89 \\ 
  46 & 2000-08-02 &   7 & -6.20 & -5.15 & -6.55 \\ 
  47 & 2000-08-05 &   5 & -8.76 & -7.42 & -8.94 \\ 
  48 & 2000-08-06 &   9 & -8.14 & -7.97 & -8.36 \\ 
  49 & 2000-08-08 &   0 & -7.99 & -7.50 & -8.11 \\ 
  50 & 2000-08-09 &   1 & -7.43 & -7.52 & -7.58 \\ 
  51 & 2000-08-10 &   9 & -9.17 & -8.11 & -7.51 \\ 
  52 & 2000-08-11 &   7 & -9.51 & -8.23 & -9.97 \\ 
  53 & 2000-08-11 &   4 & -8.67 & -7.94 & -9.81 \\ 
  54 & 2000-08-12 &   2 & -10.57 & -9.12 & -10.84 \\ 
  55 & 2000-08-13 &   9 & -9.01 & -9.27 & -10.05 \\ 
  56 & 2000-08-14 &   2 & -8.27 & -8.60 & -9.01 \\
 \hline
\end{tabular}
\end{table}


\begin{table}[ht]
\label*{table 4}
\centering
\begin{tabular}{rlrrrr}
  \hline
 S/N& Date & Hr & FD1\% & FD2\% & FD3\% \\ 
  \hline 
  57 & 2000-08-14 &   5 & -8.05 & -8.78 & -8.30 \\ 
  58 & 2000-08-15 &   6 & -7.40 & -7.72 & -8.36 \\ 
  59 & 2000-08-16 &   4 & -7.35 & -8.44 & -7.75 \\ 
  60 & 2000-08-17 &   1 & -6.70 & -8.19 & -7.59 \\ 
  61 & 2000-08-17 &   3 & -6.74 & -7.84 & -7.35 \\ 
  62 & 2000-08-19 &   0 & -5.33 & -5.71 & -5.96 \\ 
  63 & 2000-08-20 &   0 & -4.38 & -6.04 & -5.10 \\ 
  64 & 2000-08-21 &   8 & -4.94 & -5.51 & -5.47 \\ 
  65 & 2000-08-23 &   3 & -3.80 & -4.62 & -4.67 \\ 
  66 & 2000-08-23 &   3 & -3.96 & -4.25 & -4.86 \\ 
  67 & 2000-08-24 &   2 & -4.24 & -4.20 & -5.17 \\ 
  68 & 2000-08-28 &   4 & -3.40 & -4.28 & -4.40 \\ 
  69 & 2000-08-30 &   7 & -3.85 & -3.94 & -4.88 \\ 
  70 & 2000-09-01 &   5 & -4.60 & -3.57 & -3.83 \\ 
  71 & 2000-09-02 &   7 & -5.26 & -4.57 & -3.91 \\ 
  72 & 2000-09-02 &   5 & -5.82 & -3.99 & -4.51 \\ 
  73 & 2000-09-04 &   9 & -5.28 & -3.98 & -4.53 \\ 
  74 & 2000-09-05 &   5 & -5.19 & -4.11 & -4.30 \\ 
  75 & 2000-09-09 &   3 & -6.52 & -5.76 & -5.58 \\ 
  76 & 2000-09-12 &   6 & -6.06 & -4.48 & -5.27 \\ 
  77 & 2000-09-20 &   4 & -7.71 & -6.50 & -6.61 \\ 
  78 & 2000-09-21 &   1 & -7.32 & -5.67 & -6.05 \\ 
  79 & 2000-09-22 &   3 & -6.48 & -5.67 & -5.25 \\ 
  80 & 2000-09-23 &   2 & -6.14 & -5.37 & -5.53 \\ 
  81 & 2000-09-23 &   8 & -5.51 & -4.78 & -4.64 \\ 
  82 & 2000-09-23 &   2 & -5.34 & -4.40 & -4.42 \\ 
  83 & 2000-09-24 &   1 & -4.46 & -3.56 & -3.57 \\ 
  84 & 2000-09-24 &   7 & -4.14 & -3.56 & -3.89 \\ 
  85 & 2000-09-26 &   8 & -5.11 & -3.81 & -4.19 \\ 
  86 & 2000-10-02 &   5 & -4.56 & -4.11 & -4.11 \\ 
  87 & 2000-10-04 &   6 & -5.56 & -4.73 & -4.10 \\ 
  88 & 2000-10-06 &   7 & -5.50 & -3.47 & -5.15 \\ 
  89 & 2000-10-09 &   9 & -4.56 & -3.86 & -4.36 \\ 
  90 & 2000-10-09 &   3 & -3.76 & -3.73 & -3.82 \\ 
  91 & 2000-10-13 &   2 & -4.39 & -3.24 & -4.04 \\ 
  92 & 2000-10-13 &   4 & -4.55 & -3.49 & -4.04 \\ 
  93 & 2000-10-31 &   4 & -6.63 & -4.45 & -6.03 \\ 
  94 & 2000-11-01 &   4 & -6.21 & -4.87 & -4.90 \\ 
  95 & 2000-11-02 &   2 & -5.36 & -5.32 & -4.64 \\ 
  96 & 2000-11-12 &   1 & -6.66 & -5.74 & -5.87 \\ 
  97 & 2000-11-13 &   1 & -6.43 & -5.49 & -5.94 \\ 
  98 & 2000-11-18 &   3 & -6.10 & -4.31 & -4.21 \\ 
  99 & 2000-11-22 &   0 & -6.63 & -4.45 & -4.67 \\ 
  100 & 2000-11-24 &   0 & -6.04 & -4.65 & -4.90 \\ 
  101 & 2000-11-24 &   0 & -5.92 & -4.73 & -4.51 \\ 
  102 & 2000-11-30 &   0 & -11.73 & -10.41 & -10.60 \\ 
  103 & 2000-12-01 &   6 & -9.12 & -7.44 & -7.88 \\ 
  104 & 2000-12-01 &   8 & -9.11 & -8.44 & -7.71 \\ 
  105 & 2000-12-02 &   7 & -8.06 & -7.10 & -6.60 \\ 
  106 & 2000-12-04 &   0 & -7.69 & -6.60 & -6.26 \\ 
  107 & 2000-12-05 &   6 & -7.60 & -6.01 & -6.63 \\ 
  108 & 2000-12-06 &   1 & -6.65 & -6.41 & -5.41 \\ 
  109 & 2000-12-07 &   3 & -5.82 & -5.17 & -5.09 \\ 
  110 & 2000-12-09 &   6 & -5.37 & -4.06 & -4.64 \\
 \hline
\end{tabular}
\end{table}


\begin{table}[ht]
\label*{table 4}
\centering
\begin{tabular}{rlrrrr}
  \hline
 S/N& Date & Hr & FD1\% & FD2\% & FD3\% \\ 
  \hline 
 
  111 & 2000-12-12 &   3 & -4.58 & -4.40 & -4.11 \\ 
  112 & 2000-12-13 &   6 & -4.28 & -3.57 & -3.45 \\ 
  113 & 2000-12-13 &   3 & -4.42 & -3.39 & -3.79 \\ 
  114 & 2000-12-22 &   0 & -6.06 & -3.49 & -4.40 \\ 
  115 & 2000-12-22 &   2 & -6.06 & -3.91 & -4.16 \\ 
  116 & 2000-12-24 &   4 & -5.45 & -4.31 & -4.05 \\ 
  117 & 2000-12-25 &   3 & -5.76 & -5.62 & -4.65 \\ 
  118 & 2000-12-26 &   8 & -6.43 & -4.65 & -5.28 \\ 
  119 & 2000-12-31 &   7 & -4.41 & -3.31 & -3.06 \\ 
  120 & 2000-12-31 &   3 & -4.53 & -3.78 & -3.38 \\ 
  121 & 2001-01-03 &   0 & -4.95 & -4.20 & -3.29 \\ 
  122 & 2001-01-03 &   8 & -4.78 & -3.99 & -3.73 \\ 
  123 & 2001-01-07 &   2 & -4.55 & -3.02 & -3.10 \\ 
  124 & 2001-01-07 &   1 & -4.24 & -3.47 & -3.06 \\ 
  125 & 2001-01-16 &   2 & -5.33 & -3.64 & -3.33 \\ 
  126 & 2001-01-17 &   3 & -4.94 & -3.78 & -3.00 \\ 
  127 & 2001-01-26 &   6 & -6.04 & -4.36 & -4.23 \\ 
  128 & 2001-01-27 &   7 & -5.05 & -3.44 & -3.19 \\ 
  129 & 2001-04-12 &   1 & -13.77 & -13.47 & -12.78 \\ 
  130 & 2001-04-16 &   4 & -6.07 & -6.04 & -5.84 \\ 
  131 & 2001-08-23 &   6 & -3.13 & -3.39 & -3.78 \\ 
  132 & 2001-08-24 &   4 & -3.10 & -3.99 & -3.01 \\ 
  133 & 2001-08-24 &   7 & -3.87 & -3.19 & -3.45 \\ 
  134 & 2001-08-28 &   0 & -8.39 & -9.15 & -9.06 \\ 
  135 & 2001-08-30 &   0 & -7.54 & -6.51 & -6.68 \\ 
  136 & 2001-09-02 &   8 & -4.95 & -4.08 & -4.67 \\ 
  137 & 2001-09-07 &   1 & -3.37 & -3.02 & -3.94 \\ 
  138 & 2001-09-30 &   3 & -8.61 & -10.16 & -8.34 \\ 
  139 & 2001-10-05 &   5 & -6.35 & -5.24 & -4.96 \\ 
  140 & 2001-10-05 &   0 & -5.59 & -5.91 & -5.33 \\ 
  141 & 2001-10-07 &   1 & -4.49 & -4.20 & -4.46 \\ 
  142 & 2001-10-14 &   2 & -5.19 & -3.91 & -5.25 \\ 
  143 & 2001-10-21 &   9 & -4.46 & -5.24 & -3.91 \\ 
  144 & 2001-10-22 &   5 & -6.73 & -5.46 & -5.06 \\ 
  145 & 2001-10-23 &   8 & -4.61 & -3.86 & -4.06 \\ 
  146 & 2001-10-24 &   7 & -3.29 & -3.10 & -4.15 \\ 
  147 & 2001-10-26 &   8 & -3.43 & -3.69 & -3.88 \\ 
  148 & 2001-10-26 &   9 & -3.71 & -3.56 & -3.89 \\ 
  149 & 2001-10-27 &   7 & -3.93 & -3.99 & -4.84 \\ 
  150 & 2001-10-28 &   6 & -5.31 & -4.53 & -5.52 \\ 
  151 & 2001-10-29 &   1 & -3.93 & -4.62 & -3.48 \\ 
  152 & 2001-11-08 &   3 & -3.57 & -3.73 & -3.89 \\ 
  153 & 2001-12-04 &   9 & -3.48 & -3.07 & -3.22 \\ 
  154 & 2001-12-06 &   2 & -3.51 & -3.83 & -3.48 \\ 
  155 & 2001-12-07 &   7 & -3.63 & -3.69 & -3.38 \\ 
  156 & 2002-01-02 &   1 & -6.43 & -7.10 & -6.50 \\ 
  157 & 2002-01-03 &   2 & -7.69 & -7.10 & -7.39 \\ 
  158 & 2002-01-03 &   4 & -7.89 & -7.25 & -7.23 \\ 
  159 & 2002-01-05 &   6 & -6.49 & -6.01 & -5.69 \\ 
  160 & 2002-01-08 &   2 & -3.88 & -4.06 & -3.74 \\ 
  161 & 2002-01-12 &   1 & -6.84 & -6.38 & -5.91 \\ 
  162 & 2002-01-12 &   4 & -6.43 & -6.46 & -6.18 \\ 
  163 & 2002-01-12 &   6 & -6.54 & -6.41 & -6.10 \\ 
  164 & 2002-01-13 &   0 & -6.52 & -5.62 & -6.05 \\ 
  165 & 2002-01-14 &   0 & -5.59 & -6.18 & -5.86 \\ 
  166 & 2002-01-15 &   1 & -5.59 & -5.07 & -4.91 \\
 \hline
\end{tabular}
\end{table}


\begin{table}[ht]
\label*{table 4}
\centering
\begin{tabular}{rlrrrr}
  \hline
 S/N& Date & Hr & FD1\% & FD2\% & FD3\% \\ 
  \hline 
 
  167 & 2002-01-15 &   3 & -5.62 & -5.09 & -5.21 \\ 
  168 & 2002-01-16 &   5 & -4.83 & -4.33 & -4.59 \\ 
  169 & 2002-01-16 &   0 & -3.87 & -4.33 & -4.26 \\ 
  170 & 2002-01-24 &   7 & -3.52 & -3.57 & -3.33 \\ 
  171 & 2002-01-29 &   0 & -4.24 & -5.32 & -4.43 \\ 
  172 & 2002-01-29 &   9 & -5.44 & -5.46 & -5.05 \\ 
  173 & 2002-01-30 &   1 & -5.30 & -5.71 & -5.12 \\ 
  174 & 2002-02-01 &   0 & -5.70 & -5.54 & -5.26 \\ 
  175 & 2002-02-02 &   3 & -4.67 & -4.78 & -4.86 \\ 
  176 & 2002-02-02 &   8 & -3.76 & -4.57 & -4.37 \\ 
  177 & 2002-03-19 &   0 & -4.49 & -4.87 & -4.27 \\ 
  178 & 2002-03-20 &   6 & -4.66 & -4.65 & -3.56 \\ 
  179 & 2002-03-21 &   0 & -5.81 & -5.79 & -5.28 \\ 
  180 & 2002-03-24 &   7 & -6.26 & -5.40 & -4.79 \\ 
  181 & 2002-03-25 &   0 & -5.50 & -7.17 & -5.20 \\ 
  182 & 2002-03-26 &   3 & -5.39 & -4.83 & -5.52 \\ 
  183 & 2002-03-27 &   4 & -4.61 & -4.28 & -4.16 \\ 
  184 & 2002-03-28 &   2 & -4.07 & -4.08 & -3.74 \\ 
  185 & 2002-03-31 &   3 & -4.18 & -3.27 & -3.77 \\ 
  186 & 2002-04-19 &   5 & -5.09 & -3.56 & -4.07 \\ 
  187 & 2002-04-19 &   3 & -5.82 & -4.48 & -6.03 \\ 
  188 & 2002-04-20 &   3 & -5.95 & -5.66 & -7.10 \\ 
  189 & 2002-04-21 &   8 & -5.05 & -4.45 & -5.18 \\ 
  190 & 2002-04-21 &   3 & -4.58 & -4.53 & -5.37 \\ 
  191 & 2002-04-26 &   8 & -4.92 & -4.58 & -5.62 \\ 
  192 & 2002-04-27 &   2 & -4.63 & -4.48 & -4.94 \\ 
  193 & 2002-04-27 &   5 & -4.58 & -4.33 & -4.93 \\ 
  194 & 2002-05-20 &   2 & -3.87 & -4.08 & -4.75 \\ 
  195 & 2002-05-22 &   7 & -4.74 & -5.07 & -5.54 \\ 
  196 & 2002-05-23 &   2 & -7.22 & -5.76 & -7.90 \\ 
  197 & 2002-06-11 &   4 & -3.18 & -3.36 & -3.52 \\ 
  198 & 2002-06-19 &   2 & -3.65 & -3.19 & -3.30 \\ 
  199 & 2002-07-10 &   8 & -3.49 & -3.52 & -3.59 \\ 
  200 & 2002-07-22 &   8 & -4.84 & -4.92 & -5.05 \\ 
  201 & 2002-07-23 &   8 & -5.39 & -5.40 & -5.07 \\ 
  202 & 2002-07-28 &   4 & -5.70 & -7.18 & -5.31 \\ 
  203 & 2002-07-28 &   7 & -6.46 & -7.39 & -5.80 \\ 
  204 & 2002-07-28 &   9 & -6.85 & -8.19 & -7.50 \\ 
  205 & 2002-08-01 &   2 & -7.52 & -9.37 & -8.18 \\ 
  206 & 2002-08-02 &   4 & -10.04 & -9.37 & -9.65 \\ 
  207 & 2002-08-03 &   3 & -8.36 & -7.89 & -8.60 \\ 
  208 & 2002-08-05 &   4 & -7.89 & -6.26 & -6.73 \\ 
  209 & 2002-08-07 &   6 & -5.44 & -6.80 & -5.44 \\ 
  210 & 2002-08-10 &   6 & -5.20 & -5.40 & -5.33 \\ 
  211 & 2002-08-18 &   8 & -4.11 & -4.28 & -4.40 \\ 
  212 & 2002-08-19 &   7 & -6.26 & -6.85 & -4.22 \\ 
  213 & 2002-08-20 &   5 & -7.52 & -6.63 & -7.35 \\ 
  214 & 2002-08-21 &   6 & -6.28 & -5.84 & -6.82 \\ 
  215 & 2002-08-22 &   3 & -7.01 & -6.18 & -6.61 \\ 
  216 & 2002-08-24 &   8 & -7.10 & -6.46 & -6.22 \\ 
  217 & 2002-08-28 &   0 & -8.23 & -8.60 & -7.63 \\ 
  218 & 2002-08-31 &   4 & -6.14 & -6.46 & -6.08 \\ 
  219 & 2002-09-02 &   1 & -4.78 & -5.00 & -5.00 \\ 
  220 & 2002-09-04 &   5 & -6.21 & -5.54 & -5.63 \\ 
  221 & 2002-09-07 &   5 & -4.63 & -5.17 & -3.78 \\ 
  222 & 2002-09-09 &   4 & -5.34 & -5.96 & -5.80 \\
\hline
\end{tabular}
\end{table}


\begin{table}[ht]
\label*{table 4}
\centering
\begin{tabular}{rlrrrr}
  \hline
 S/N& Date & Hr & FD1\% & FD2\% & FD3\% \\ 
  \hline 
 
  223 & 2002-09-10 &   1 & -5.84 & -4.78 & -5.15 \\ 
  224 & 2002-09-13 &   9 & -4.55 & -4.20 & -4.10 \\ 
  225 & 2002-09-21 &   0 & -3.41 & -4.41 & -4.10 \\ 
  226 & 2002-09-21 &   7 & -4.74 & -3.94 & -4.93 \\ 
  227 & 2002-09-22 &   5 & -4.39 & -5.12 & -3.75 \\ 
  228 & 2002-09-24 &   7 & -5.95 & -4.28 & -5.27 \\ 
  229 & 2002-09-25 &   1 & -3.93 & -4.70 & -3.98 \\ 
  230 & 2002-09-27 &   9 & -4.74 & -4.20 & -4.22 \\ 
  231 & 2002-09-28 &   3 & -4.89 & -4.50 & -3.75 \\ 
  232 & 2002-09-30 &   1 & -3.63 & -3.61 & -3.62 \\ 
  233 & 2002-10-04 &   5 & -4.88 & -4.50 & -3.91 \\ 
  234 & 2002-10-19 &   4 & -5.67 & -4.40 & -4.44 \\ 
  235 & 2002-10-20 &   6 & -5.75 & -5.76 & -4.88 \\ 
  236 & 2002-10-22 &   6 & -6.24 & -6.13 & -5.22 \\ 
  237 & 2002-10-23 &   6 & -5.08 & -5.15 & -5.02 \\ 
  238 & 2002-10-26 &   2 & -5.12 & -5.62 & -3.98 \\ 
  239 & 2002-10-30 &   5 & -3.79 & -3.81 & -3.16 \\ 
  240 & 2002-11-02 &   3 & -5.08 & -5.74 & -5.05 \\ 
  241 & 2002-11-06 &   4 & -5.51 & -5.76 & -5.74 \\ 
  242 & 2002-11-08 &   0 & -3.15 & -4.06 & -3.56 \\ 
  243 & 2002-11-08 &   4 & -3.48 & -3.44 & -3.47 \\ 
  244 & 2002-11-11 &   1 & -4.24 & -5.24 & -5.63 \\ 
  245 & 2002-11-14 &   2 & -4.47 & -6.60 & -4.78 \\ 
  246 & 2002-11-16 &   7 & -3.68 & -4.65 & -3.62 \\ 
  247 & 2002-11-19 &   3 & -7.92 & -8.81 & -8.56 \\ 
  248 & 2002-11-22 &   9 & -4.81 & -5.37 & -5.48 \\ 
  249 & 2002-11-23 &   2 & -3.94 & -5.40 & -4.30 \\ 
  250 & 2002-11-25 &   9 & -3.69 & -4.65 & -4.00 \\ 
  251 & 2002-11-27 &   2 & -5.06 & -5.62 & -4.65 \\ 
  252 & 2002-11-28 &   0 & -5.03 & -5.79 & -4.58 \\ 
  253 & 2002-11-29 &   9 & -4.14 & -3.91 & -3.98 \\ 
  254 & 2002-11-30 &   6 & -3.18 & -3.99 & -3.73 \\ 
  255 & 2002-12-09 &   7 & -5.02 & -5.20 & -4.57 \\ 
  256 & 2002-12-10 &   8 & -4.42 & -5.00 & -4.59 \\ 
  257 & 2002-12-12 &   2 & -3.77 & -4.36 & -3.56 \\ 
  258 & 2002-12-16 &   0 & -5.14 & -5.49 & -4.52 \\ 
  259 & 2002-12-24 &   6 & -6.99 & -7.02 & -5.89 \\ 
  260 & 2002-12-24 &   9 & -6.80 & -6.35 & -6.20 \\ 
  261 & 2002-12-28 &   3 & -4.02 & -5.84 & -3.50 \\ 
   \hline
\end{tabular}
\end{table}
\section {Acknowledgments}

We gratefully acknowledge the team that hosts the website {http://cr0.izmiran.rssi.ru/ and https://omniweb.gsfc.nasa.gov/html/ow data.html} from where we obtained the data for this work at no cost. Of special note  is the non-commercial R software  developers and  the members of the R-mailing list (R-help@r-project.org). To you all, we are sincerely indebted.  The current manuscript has enjoyed the benefit of meeting an expert in the field. The manuscript was significantly changed following the recommendations of the unknown reviewer. We feel indebted to the editor for sending our manuscript to him/her. Our friend, Jim Lemon, has also added his expertise both in proofreading, editing and data analysis. His efforts are duly acknowledged. Finally, JAA wish to appreciate the leadership of Astronomy and Astrophysics Research Lab group, University of Nigeria, Nsukka for sound academic leadership.

\bibliography{reference}

\end{document}